\documentclass[]{article}
\usepackage[skip=0pt]{parskip}
\usepackage{arxiv}
\usepackage[utf8]{inputenc} 
\usepackage[T1]{fontenc}    
\usepackage{hyperref}       
\usepackage{url}            
\usepackage{booktabs}       
\usepackage{amsfonts}       
\usepackage{nicefrac}       
\usepackage{microtype}      
\usepackage{lipsum}
\usepackage{tcolorbox}
\usepackage{graphicx}
\usepackage{fancyhdr}
\usepackage{natbib}
\usepackage[affil-it]{authblk}
\usepackage{index, amsmath, listings, multirow, nomencl, ltablex, longtable,  rotating, scalerel,  setspace, subfig,tabularx, titlesec,url, verbatim, adjustbox, algpseudocode, algorithmicx, algorithm, comment}
\usepackage[table,xcdraw]{xcolor}

\makeatletter
\renewcommand{\normalsize}{%
  \@setfontsize\normalsize{12pt}{12pt}
  \abovedisplayskip      8\p@ \@plus 2\p@ \@minus 5\p@
  \abovedisplayshortskip \z@ \@plus 3\p@
  \belowdisplayskip      \abovedisplayskip
  \belowdisplayshortskip 4\p@ \@plus 3\p@ \@minus 3\p@
}

\normalsize
\makeatother

\graphicspath{Figures/}
\title{Capturing the Curve: Functional Data Analysis for Validated Digital Outcome Measures}
\author[1*]{Mia S. Tackney}
\author[2]{Marcos Matabuena}
\author[3]{Marco Palma}
\author[4]{Michael Wester}
\author[4]{Claire Maa\ss en}
\author[4]{Thomas Krammer}
\author[5, 4]{Julian Mustroph}
\author[6]{Peter H. Charlton}
\author[7, 8]{James Carpenter}
\author[1, 8]{Sofía S. Villar}
\affil[*]{Corresponding author, \texttt{mst35@cam.ac.uk}}
\affil[1]{MRC-Biostatistics Unit, University of Cambridge, UK}
\affil[2]{Mohamed bin Zayed University of Artificial Intelligence, Abu Dhabi, UAE}
\affil[3]{Population, Policy and Practice Research and Teaching Department,
 UCL Great Ormond Street Institute of Child Health, London, UK }
\affil[4]{University Heart Center Regensburg, University of Regensburg, Regensburg, Germany}
\affil[5]{Department of Pharmacology, University of Regensburg, Regensburg, Germany}
\affil[6]{Department of Public Health and Primary Care, University of Cambridge, Cambridge,
UK}
\affil[7]{Department of Medical Statistics, London School of Hygiene and Tropical Medicine, London, UK}
\affil[8]{Medical Research Council Centre of Research Excellence in Clinical Trial Innovation (CCTI), Institute of Clinical Trials and Methodology, UCL, London, UK}
\begin{document}
\maketitle
\begin{abstract}
Digital health devices and other passive monitoring technologies enable high-frequency collection of health outcomes in near-continuous time, with the potential to capture rich information about the health of individuals. The raw data collected by these devices often have a hierarchical functional structure: repeated physiological functions are observed over time and on multiple time scales (seconds, days, weeks). In clinical trials, while many summaries can be derived from digital data, typically, only a small subset of pre-defined scalars are validated as clinical outcome measures. This paper proposes a Multilevel Functional Principal Component Analysis (MFPCA) framework to obtain low-dimensional representations of functional data with robust statistical properties, which has received limited attention in clinical trial settings. Specifically, we compute MFPCA projection scores with respect to a reference (healthy or control) population and use subject-level scores as the summary metric, which quantify how individuals differ from the dominant subject-level directions of variation. Through a simulation study based on smartwatch electrocardiogram (ECG) signals, we compare MFPCA scores with pre-specified summaries in terms of validation criteria for digital outcome measures, including test–retest reliability and known-groups discrimination. We demonstrate that MFPCA scores generally have high reliability, and across simulated scenarios of change between two groups, at least one MFCPA score can discriminate between groups. This offers an advantage when digital tools enable the measurement of novel physiological signals and the characteristics of change are not yet defined. Finally, using knee flexion–extension data from individuals living with Parkinson’s disease, we demonstrate that one of the MFPCA scores  correlated more strongly with established gold-standard metrics and can detect clinical change, compared to a pre-specified scalar. We conclude that MFPCA-derived scores offer a promising framework for developing digital outcomes that retain more information than many classical outcome measures and open the door to using learning representation strategies in clinical trial settings.
\end{abstract}
\keywords{Digital Outcome Measures \and Functional Data Analysis  \and Clinical Validation  \and Functional Multilevel Principal Components Analysis \and Medical device \and Smartwatch \and Apple Watch}
\section{Background}
\begin{figure}[H]
\centering
\includegraphics[width=1\textwidth]{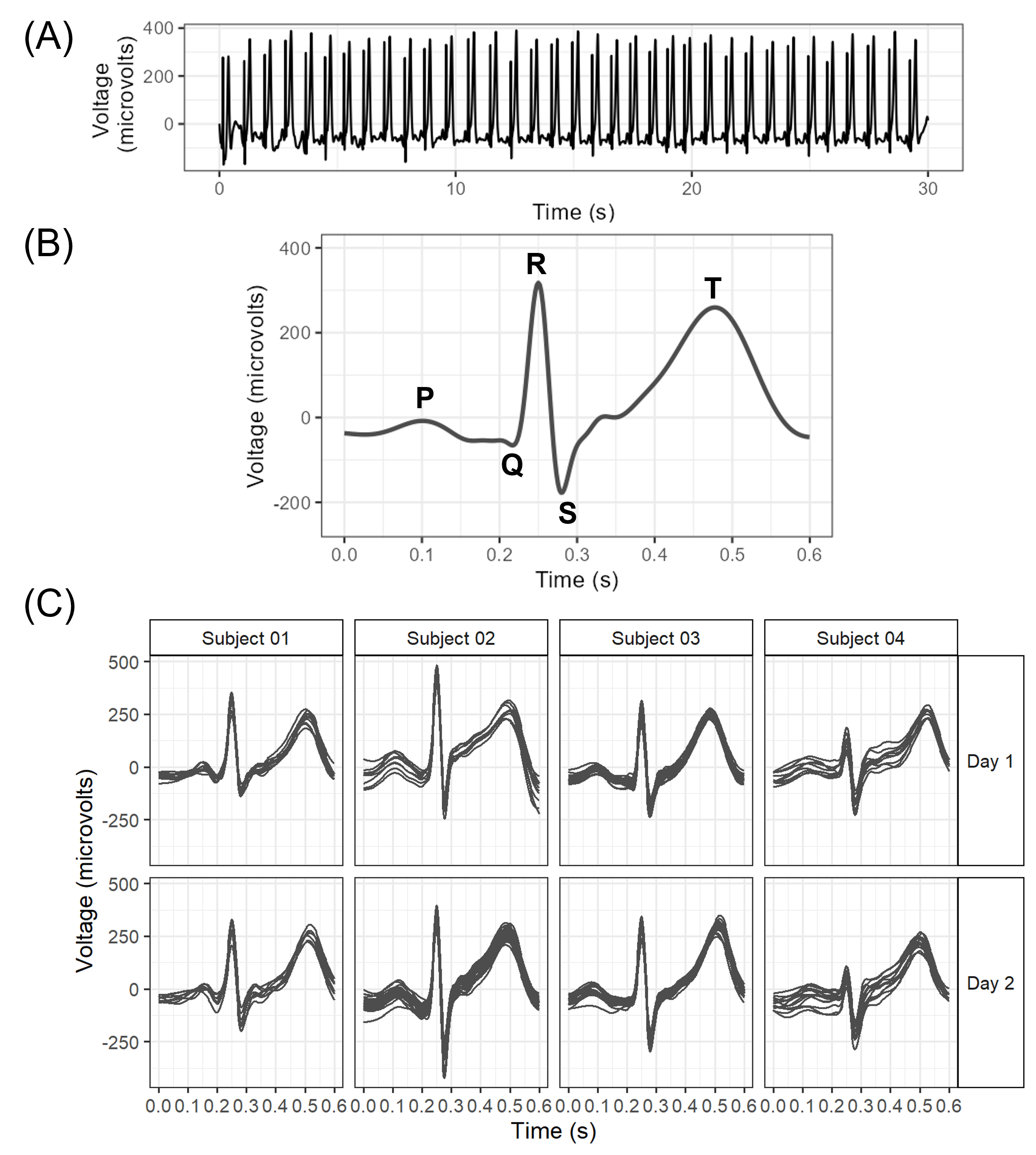}
\caption{Apple watch ECGs. (A) a 30-second Apple watch ECG recording from an individual; a single ECG cardiac cycle, with features of clinical interest indicated: the P wave represents atrial depolarization, the QRS complex (Q, R, and S waves) represents ventricular depolarization, and the T wave represents ventricular repolarization; (B) The hierarchical nature of AW ECG curves is illustrated. Firstly, there are four individuals. Secondly, there are ECG recordings from two days per individual. Thirdly, for each recording, there are several repeated cardiac cycles.}
\label{example_ECGs}
\end{figure}

Digital technology enables measurements of physiological variables at high frequencies; for example, strides can be measured via accelerometers at a high frequency over periods ranging from days to weeks, and electrocardiogram (ECG) data can be collected from smartwatches through convenient 30-second tests. While these technologies enable remote monitoring and potential improvements in data collection for trials and clinical care, the volume of data that arises, and their complex hierarchical structure, pose challenges for data analysis \citep{Di2009}. As an illustration, we describe Apple Watch (AW) ECG data which serves as a running example in this article. Figure \ref{example_ECGs} panel (A) displays a 30-second ECG recording. Panel (B) displays a single cardiac cycle with the key waveforms highlighted. Panel (C) illustrates the hierarchical nature of AW ECG data: we have data from four individuals, representing the first level of hierarchy. ECGs from two days are shown for each individual, representing the second level of hierarchy. Lastly, in each 30 ECG-second reading, many cardiac cycles are obtained, illustrating the third level of hierarchy. We can assume that the curves are observations from a smooth process evolving over a temporal functional domain. We observe differences in individuals not only in terms of timing and amplitudes of peaks, but waveform shape, symmetry (and asymmetry) and slopes, illustrating the advantages of analysing the entire curve. Furthermore, we observe variability within individuals; there is greater variability in the within-occasion curves of Subject 04 compared to Subject 03. Decomposing different sources of variability is crucial in this setting, as it can help identify which sources of variability which are clinically meaningful.  \\

Given that high-frequency measuresments from digital technologies may capture physiological measurements precisely and conveniently in individual's daily environments, there is interest in their use to measure health outcomes in clinical trials \citep{FoodandDrugAdministration2023}. These devices can capture fluctuations and trends over longer periods that may be missed via traditional trial outcomes, which typically are collected in episodic clinical visits, such as the 6 Minute Walk Test (6MWT)  \citep{Landers2021, Tackney2024}. However, despite enthusiasm for their use, there remain critical gaps in how this large volume of data should be analysed. Typical practice is to summarise the data into a scalar metric to represent the individual's health outcome at a specific time point in the trial (e.g. baseline, 3-month follow-up or 6-month follow-up), which leads to loss of information about the trajectory of the outcome over that period. Using pre-specified scalars has historically been a prerequisite for validating digital outcomes for regulatory qualification; for example, a landmark case is the European Medicines Agency’s (EMA) qualification of the Stride Velocity 95th Centile (SV95C) for Duchenne Muscular Dystrophy \citep{Servais2022, Colloud2023}. Here, the SV95C was found to have superior performance in validation criteria compared to other summaries, such as median distance walked or median velocity and there was limited exploration of the functional trajectories from the digital device. \\

This article investigates alternative analytical frameworks for summarising data from digital health technologies by treating the data as functional trajectories. Because many physiological processes are inherently continuous, the standard practice of collapsing multilevel data into pre-specified scalar summaries risks discarding important signals. The shape of these repeated functions and their intra-individual variability are often clinically meaningful; intra-individual variability is lost when averaging summaries for curves within person. Consequently, we shift our focus away from fixed summary statistics and toward multilevel functional data analysis to derive data-driven metrics.\\

Functional data analysis has been used to analyse sensor data, particularly in observational settings; see, for example \citep{Zhou2024, Goldsmith2015, Matabuena2023} for accelerometer data, \citep{Di2009} for electroencephalographic data (EEG), \citep{Matabuena2026, Yang2024}) for continuous glucose monitor data (CGM) and \citep{Minhas2025, Roach2021, Yoshida2022} for gait data. Despite its advantages, the application of functional data analysis within clinical trial settings remains very limited. This may be driven by the need to strictly pre-specify outcome measures in Statistical Analysis Plans (SAPs), a general preference for scalar metrics with established clinical interpretability, and high thresholds for regulatory acceptance, particularly when defininng a primary outcome measure. By utilising multilevel functional data analysis to derive data-driven summaries for clinical validation of digital outcome measures, this work demonstrates how functional approaches can be integrated into the clinical trials setting.   \\

Through a simulation study motivated by smartwatch ECG outcomes, we demonstrate how data-driven scalars from a multilevel functional principal components analysis (MFPCA) \citep{Di2009}, computed with respect to a reference (healthy) dataset, can quantify how much an individual deviates from the reference (healthy) distribution. We illustrate how projection scores can be obtained for new data based on the MFCPA for the reference dataset, which summarise the deviation of new individuals from the dominant modes of variation in the reference group. We compare performance of typical summary metrics (such as amplitudes of peaks of the ECG) alongside scores from an MFPCA on two important criteria for clinical validation: reliability and known-groups discrimination. We then analyse knee flexion/extension data from individuals living with Parkinson's disease using MFPCA and compare the correlation of these scores with the gold-standard MDS-UPDRS Part II and III scores at a single measurement occasion, as well as in its change across two measurement occasions, one where individuals are on medication and one where individuals are off medication.\\

The article is structured as follows. In Section \ref{Validation_criteria}, we describe the criteria for validation of digital outcome measures. In Section \ref{ECG}, we introduce the motivating example of ECG data from smartwatches and take note of specific complications in this setting, including the need for landmark registration. In Section \ref{functional}, using the example of smartwatch ECG data, we define the notation and framework for MFPCA. We illustrate how, after conducting an MFPCA on a reference dataset, projected scores from a new dataset onto the reference dataset can be obtained, which reflect how each individual's curves deviate with respect to the reference dataset. The simulation study in Section \ref{Sim} compares performance of pre-specified summaries against scores from an MPFCA in terms of intra-cluster correlation and ability to discriminate between groups. Section \ref{PDdata} introduces gait data from individuals living with Parkinson's Disease, as well as healthy individuals, and illustrates how scores from an MFPCA compare against a scalar summary in terms of concurrent validity and ability to detect change. Finally, in Section \ref{Discussion}, we describe the potential role for MFPCA in analyses of digital outcome data, in particular in identifying a validated outcome measure.

\section{Validation criteria}
\label{Validation_criteria}

The V3+ framework \citep{Bakker2024} (formerly known as the V3 framework \citep{Goldsack2020}) is a well-known validation framework for digital outcome measures and provides core qualities, including statistical properties, that should be targeted when validating a digital outcome measure. This framework separates validation into four key areas: (i) usability validation, (ii) verification, (iii) analytical validation and (iv) clinical validation. We focus on clinical validation, which examines whether the digital outcome can capture a clinical experience in the intended context. We describe the five clinical validation criteria and how each criterion was demonstrated for  SV95C for Duchenne Muscular Dystrophy (DMD) \citep{Servais2021} in Table \ref{sv95c_validation}. \\

\begin{itemize}
\item \textbf{Convergent validity} refers to the correlation between the gold-standard outcome and the digital outcome measure, which shows that the digital outcome measure can reliably predict the gold-standard outcome. Pearson's and Spearman's correlations may be computed between the traditional and digital outcomes. \\

\item \textbf{Known-groups validity} describes how well a metric can discriminate between groups; this could be between a group with the disease condition and a healthy control group, or between groups with different severities of disease. An appropriate test of the summary metric between the two groups may be used. \\

\item \textbf{Reliability} refers to the consistency of repeated readings. We describe two intra-cluster correlation coefficients (ICC) which are commonly reported metrics for test-retest reliability \citep{ShroutFleiss1979, McGrawWong1996, Trevethan2017, Liljequist2019}. We denote by $y_{ijk}$ the outcome for individual $i, i \in \left\{1, 2, ..., N \right\}$, on measurement occasion $j, j \in \left\{1, 2, ..., n_i \right\}$, and measurement $k, k \in \left\{1, 2, ..., m \right\}$. \cite{Ratitch2023} describe a two-way random effects analysis of variance (ANOVA) to decompose the sources of variation from a digital outcome:  

\begin{equation}
\label{two-way_ANOVA}
    y_{ijk}=\mu + r_i + c_j +v_{ijk},
\end{equation}

$\mu$ is the overall mean, $r_i\sim N(0, \sigma_r^2)$ is a random effect for subject $i$, $c_j \sim N(0, \sigma^2_c)$ is a random effect for occasion $j$ and $v_{ijk} \sim N(0, \sigma^2_v)$ is a measurement error. \\

 The ICC metrics provide a measure of signal-to-noise and signal the proportion of variation due to the individual over the total variation. ICC(A,1) is a measure of absolute agreement (where ``A'' refers to agreement) and assesses whether repeated measurements give exactly the same outcomes. Here, $c_j$ in Equation \eqref{two-way_ANOVA} is assumed to be random: 
\begin{equation*}
   \mbox{ICC}(A,1) =\frac{\sigma^2_r}{\sigma^2_r+\sigma^2_c+\sigma^2_v}.
\end{equation*}

In situations where systematic biases are considered acceptable and it is sufficient that the measurements lead to a consistent ranking order of the individuals, the consistency measure ICC(C,1), (where ``C'' refers to consistency) can be used: 
\begin{equation}
   \mbox{ICC}(C,1) =\frac{\sigma^2_r}{\sigma^2_r+\sigma^2_v}.
\end{equation}

The population ICC metrics are estimated via sample variances $\hat{\sigma}^2_r$, $\hat{\sigma}^2_c$ and $\hat{\sigma}^2_c$, calculated from the multilevel model in Equation \eqref{two-way_ANOVA}. In some situations, $c_j$ may be assumed to be fixed \citep{Ratitch2023}; for example, if there is a learning effect that systematically leads to lower values for early measurement occasions. \\

\item \textbf{Responsiveness} refers to the ability of the digital outcome measure to detect change over time, either due to disease progression or treatment. This may be demonstrated by an earlier statistically significant change from baseline due to disease progression or initiation of treatment.\\

\item \textbf{Meaningful Change} refers to whether the digital outcome measure can detect change that is defined as being meaningful, either via clinician- or patient-reported references of change (anchor-based estimates), or via statistical properties of the measure (distribution-based estimates).\\

\end{itemize}

\begin{table}[]
\caption{Criteria for clinical validation and how they were demonstrated for SV95C for DMD \citep{Servais2021}}
\centering
\renewcommand{\arraystretch}{1.3}
\rowcolors{2}{gray!10}{white} 
\begin{tabularx}{\textwidth}{p{6.5cm} X}
\toprule
\rowcolor{gray!25}
\textbf{Clinical Validation Criterion} & \textbf{How Validity Was Demonstrated for SV95C} \\
\midrule
\textbf{Convergent validity} \newline (correlation with gold-standard endpoints) & 
Pearson's and Spearman's correlations between SV95C and gold standard endpoints (6MWD, NSAA, and 4SC) at baseline and months 3, 6, 9, and 12 were shown to be significant in DMD patients. \\
\textbf{Test-retest reliability} \newline (consistency of repeated readings) & 
Intraclass correlation coefficient for SV95C in DMD patients based on two successive readings collected 1 month apart was 0.97 ($95\%$ CI: [0.947, 0.984]). ICCs stratified by age group were also high. \\
\textbf{Known-groups validity} \newline (differentiating DMD patients from controls) & 
DMD Patients had lower median SV95C values (1.563 m/s; N = 125) than controls (2.713 m/s; N = 66; $p < 0.001$ from a Mann–Whitney-U Test). 
\\
\textbf{Responsiveness} \newline (ability to detect clinical change) & 
Median relative change in SV95C and gold-standard endpoints from baseline were calculated at 3, 6, 9, and 12 months for DMD patients on a stable corticosteroid regimen. Median relative change from baseline was $- 12.842\%$ at 12 months ($p = 0.0003$ for one-sample Wilcoxon signed rank test, $N = 34$). SV95C was shown to signal change earlier than other endpoints. \\
\textbf{Meaningful change}\newline (ability to detect change which is meaningful) & 
Meaningful change analyses suggested that changes in SV95C of at least $-0.10$ m/s in DMD patients is beyond measurement error, while a -0.10 to -0.20 m/s change is meaningful. These changes were observed within 9 months in steroid-treated patients (negative) and within 6 months after steroid initiation (positive).\\
\bottomrule
\end{tabularx}
\label{sv95c_validation}
\end{table}

In the validation of a digital outcome measure for DMD, a number of pre-specified summary metrics, including mean distance walked, and median and 95th centile of Stride Velocity, were evaluated in terms of the above criteria. To maximise the utility of functional data from digital technologies, we evaluate data-driven summary metrics obtained via multilevel FPCA in terms of the validation criteria, to assess whether this is a viable  and more robust approach to obtaining summary metrics from digital health technologies.  

\section{Exemplar 1: Smartwatch ECGs}
\label{ECG}
Our motivating example is Apple Watch (AW) ECG data from healthy participants from a registry study which aimed to assess the validity of smartwatch-based ECG recordings. We note that the study is not intended to formally validate AW ECGs as digital biomarkers or trial outcomes; we are simply using the data as an illustrative example to demonstrate statistical methods in the validation of digital outcome measures. \\

Smartwatches allow participants to conveniently record electrocardiograms (ECGs) over multiple occasions across time, provide an opportunity to monitor cardiovascular health, and may enable detection of cardiovascular disease \citep{Nazarian2021, Isakadze2020}. Smartwatch ECG recordings are typically 30 seconds long and sampled at a high frequency, such as 512Hz (512 measurements per second). As shown in Figure \ref{example_ECGs} (B), the 30-second interval consists of multiple consecutive cardiac cycles characterised by waveforms including the P wave, QRS complex, and T wave. Data from smartwatch ECGs are functions in a nested, hierarchical structure with multiple levels of variation (see Figure \ref{example_ECGs} (C)).\\

In this study, patients provided an AW ECG and a standard 12-channel ECG during an in-clinic assessment. Healthy participants were defined as inpatients who did not have cardiac disease (e.g. pre-operative assessment of orthopedic patients). Patients with cardiac disease were defined as having any cardiac disease in their medical history. Symptoms and medical history were routinely assessed by experienced physicians. We included 59 healthy patients in the analysis. Mean age was 55.5 years, 42.4\% of patients were were female, and mean BMI was 27.8 $kg/m^2$.\\

Typical summary metrics of interest in ECGs are amplitudes of peaks such as the P‑, R‑ and T‑waves. Changes in these amplitudes can reflect important clinical conditions. For example, in atrial fibrillation, absence of distinct P‑waves is a key feature \citep{Censi2016}; in cardiac amyloidosis, low overall ECG voltages can lead to reduced amplitudes across all peaks \citep{Cipriani2022}. Furthermore, widths of these waves, and intervals between specific points of curve (e.g. PT-intervals) are common summary metrics used for ECGs. \\

In our investigation, we analyse the ECG curve as a continuous function via functional data analysis to avoid information loss associated with using discrete scalar summaries like amplitudes and intervals. However, we note a specific challenge of using functional approaches in the ECG setting. ECG signals inherently exhibit variability in both amplitude (voltage) and phase (timing). Because variation in phase can obscure true differences in amplitude between curves, we isolate amplitude variations by applying landmark registration \citep[p. 132]{RamsaySilverman2005}. This process transforms the time domain so that the positions of key landmarks (in our case, P, R, and T-peaks) occur at the same time for all cardiac cycles. To achieve this, we retain only the ECG curves have all three peaks reliably detected within plausible ranges for analysis. Because this temporal alignment intentionally removes phase variability to focus on the shape of the cardiac curve, we do not evaluate intervals (e.g., PR or QT intervals) as summary metrics in this framework, despite their established clinical importance. See Supplementary File A for a description of the data preparation process.

\section{Multilevel Functional Principal Components Analysis}
\label{functional}

We now introduce the notation and general framework for multilevel functional principal components analysis (MFPCA), using data from the AW ECGs as an illustrative example. For simplicity, we describe two-level MFCPA, where each subject provides a single ECG recording, which is subsequently segmented into repeated functions representing individual cardiac cycles. We note that it can be extended to additional hierarchical levels, as demonstrated in the  3-level models in \citep{Di2009, Matabuena2023}.  \\

Let $y_{ij}(t)$ denote the $j$th curve for individual $i$, where $i \in \{1, \dots, N\}$ and $j \in \{1, \dots, n_i\}$, allowing for a variable number of curves per individual. For time $t$ within the functional temporal domain $\mathcal{T} = [0, T]$, the function $y_{ij}(t)$ can be decomposed as follows:

\begin{equation}
    y_{ij}(t)= \mu(t) + b_i(t) + w_{ij}(t) + \epsilon_{ij}(t),
\end{equation}

\noindent where $\mu(t)$ is the overall mean function, $b_i(t)$ is the subject-level deviation from the overall mean function, and $w_{ij}(t)$ is the deviation at the curve-level, which represents the deviation between each curve and the subject-specific mean function. We assume that $b_i(t)$ and $w_{ij}(t)$ have mean zero and are mutually uncorrelated stochastic processes. The error term $\epsilon_{ij}(t)$ is also considered a stochastic process with mean zero and variance $\sigma^2$ \citep{Di2014}.\\

As $b_i(t)$ and $w_{ij}(t)$ are random processes, their structure is characterised by covariance operators, which define how points on the curve at time $t$ correlate with points at time $s$ for $s, t \in \mathcal{T}$ \citep{RamsaySilverman2005}. Multilevel functional principal component analysis (MFPCA) utilises the Karhunen-Loeve expansion to decompose the covariance operators as a linear combination of fixed basis functions \citep{Karhunen1947, Loeve1945}:

\begin{align}
    b_{i}(t)&= \sum_{k_1=1}^{\infty} \eta_{ik_1} \phi_{k_1}^{(1)} (t),\\
     w_{ij}(t)&= \sum_{k_2=1}^{\infty} \xi_{ijk_2} \phi_{k_2}^{(2)} (t),
\end{align}

\noindent where $\phi_{k_1}^{(1)}(t)$, $k_1 \in \left\{ 1, 2, ..., \infty \right\}$ and $\phi_{k_2}^{(2)} (t)$, $k_2 \in \left\{ 1, 2, ..., \infty \right\}$  are the eigenfunctions of the covariance operators corresponding to between-subject and within-subject modes of variation, respectively. The between-subject eigenfunctions capture dominant patterns that distinguish individuals from one another in their overall trajectories. The within-subject eigenfunctions represent systematic temporal deviations around each individual’s mean trajectory, reflecting intra-individual variability. The eigenfunctions are orthogonal and ordered by the proportion of variance explained. The scalars $\eta_{ik_1}$ and $\xi_{ijk_2}$ represent the between-subject and within-subject principal component scores, respectively. These are zero-mean random variables with variances $\lambda_{k_1}^{(1)}$ and $\lambda_{k_2}^{(2)}$, which quantify the magnitude of variation along their corresponding eigenfunctions. Furthermore, these scores are assumed to be uncorrelated across all indices $k_1$ and $k_2$. \\

In practice, these infinite expansions are truncated to the first $K_1$ and $K_2$ principal components to provide a parsimonious, finite-dimensional approximation of the processes which captures the majority of the functional variation. Thus, the MFPCA model is given by: 
\begin{equation}
\label{fpca}
    y_{ij}(t)= \mu(t) + \sum_{k_1=1}^{K_1} \eta_{ik_1} \phi_{k_1}^{(1)} (t) + \sum_{k_2=1}^{K_2} \xi_{ijk_2} \phi_{k_2}^{(2)} (t) + e_{ij}(t).
\end{equation}

\begin{figure}[ht!]
\centering
\includegraphics[width=1\textwidth]{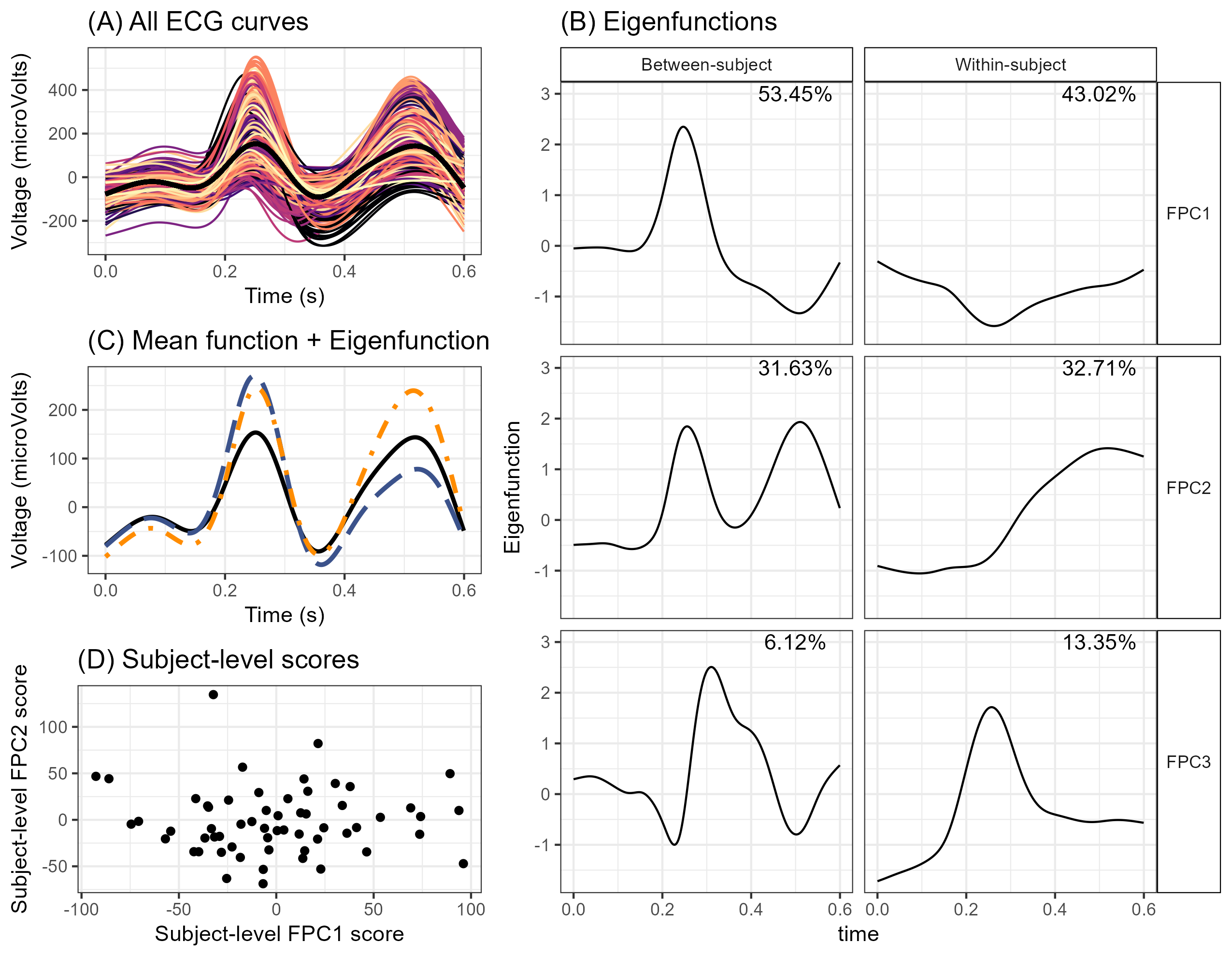}
\caption{Results from a two-way MFPCA performed on Apple Watch ECGs from 59 healthy participants: (A) Landmarked ECGs with colours distinguishing individuals. (B) The first three between- and within-subject eigenfunctions. The proportion of variance explained by each FPC over FPCs in the same level is given. (C) The overall mean function (black solid line); the mean function plus a positive multiple of the first between-subject eigenfunction (blue dotted line); the mean function plus a positive multiple of the second between-subject eigenfunction(orange dashed line). (D) Subject-level FPC2 scores plotted against FPC1 scores.}
\label{initial_fpca}
\end{figure}

\noindent We illustrate the MFPCA applied to ECG curves in Figure \ref{initial_fpca}. Panel (A) displays landmarked ECG curves from 59 healthy individuals; curves of the same colour belong to the same subject. In Panel (C), the black line illustrates the overall mean function, $\mu(t)$. Panel (B) illustrates the first three between-subject and within-subject eigenfunctions, which together capture 91.2\% and 89.3\% of the total variability at each level, respectively. The first between-subject eigenfunction appears to capture differences between the amplitudes of the R- and T-peaks between individuals, and accounts for $53.45\%$ of the between-subject variability. This may reflect that the R- and T-peaks are a dominant characteristic and may be associated with variables such as sex and height. In panel (C), an ECG curve of an individual with a high between-subject FPC1 score is shown in the dotted blue line; this leads to a higher R-peak and lower T-peak than the overall mean curve. In panel (B), the second between-subject eigenfunction has high values where the R- and T-wave occur, and explain $25.38\%$ of the between-subject variation. In panel (C), an ECG curve of an individual with a high between-subject FPC2 score is shown by the dashed orange line; this leads to higher R- and T-peaks than the overall mean curve. The third between-subject eigenfunction captures a smaller proportions of variability. The within-subject eigenfunctions represent systematic temporal deviations around each individual’s mean trajectory, reflecting intra-individual variability. Finally, in panel (D), we illustrate the low-dimensional between-subject FPC1 and FPC2 scores, which are the deviation of each individual in the direction of the first and second eigenfunctions. \\

In particular, we are interested in the between-subject scores $\eta_{ik_1}$. These are data-driven summary scores at the subject level. Unlike scalar summaries such as the P, R and T-peaks which are localised at specific points of the curve, these scores efficiently  summarise information across the entire curve into a low-dimensional scalar. \\

We note that the well-known FPCA is a special case of MFPCA where $K_2=0$ \citep{RamsaySilverman2005}, and has been used in physiological and biomechanical settings by first obtaining average curves per individual \citep{Minhas2025}. In Figure 1 in the Supplementary File A, we illustrate results of an FPCA on the 59 healthy individuals where ECG curves are first averaged to create a single function per individual. 

\subsection{Estimation of Multilevel Functional Principal Component Scores}
\label{Estimation_scores}

We described how MFPCA extracts principal modes of variation (eigenfunctions) from  physiological curve data. The principal component scores from an MFPCA are low-dimensional summaries that capture the between-subject deviations from the overall mean. However, there are challenges to how this can be applied to a clinical trial setting. Outcome measures from trials should be comparable across stages (e.g., as a multi-stage trial progresses from Stage 1 to Stage 2) and across independent trials that use the same outcome measure. Fitting MFPCA models for each stage, and for each trial, will result in inconsistent eigenfunctions and scores that are incomparable across trial stages or across independent trials.\\

\begin{figure}[ht!]
\centering
\includegraphics[width=0.5\textwidth]{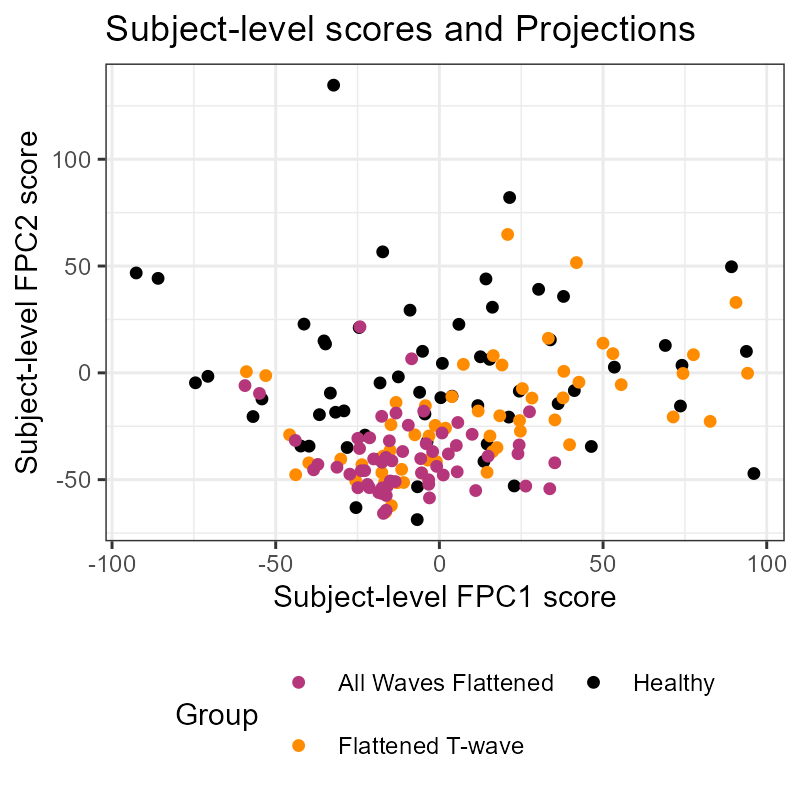}
\caption{Illustration of MFPCA scores and projections. In black, subject-level FPC2 scores are plotted against FPC1 scores from the AW ECGs from 59 healthy individuals. This groups serves as a reference group. In orange, we plot the projected between-individual scores of AW ECGs simulated to have flattened T-waves. In magenta, we plot the projected between-subject projection scores of simulated AW ECGs curves where all waves have been flattened.}
\label{plot_projections}
\end{figure}

To address this, we adopt a normative modelling framework. Rather than deriving new eigenfunctions for every dataset, we define a fixed set of eigenfunctions from a reference population. Data from healthy controls in validation studies may be a natural source for this data, such as a study of 91 healthy volunteers that provided normative values for SV95C \citep{Poleur2021}. Data from a large-scale population studies such as the UK Biobank, or from a pilot cohort, may also be used as normative data. Clinical trial data are then projected onto this established reference, to obtain between-subject scores that quantify deviations relative to the reference population. This approach ensures that the scores are comparable across stages and studies. As an illustration, suppose that the data AW ECG data from the 59 healthy participants serve as a reference group. Figure \ref{plot_projections} displays in black the FPC2 scores against the FPC1 scores from this group. Next, suppose that we have data from individuals whose T-waves are flattened. Their projected scores, displayed in orange, show that the distribution of scores have changed slightly; fewer individuals have low values of FPC1, and fewer individuals have high values of FPC2. Further, in magenta, we plot the projected scores of individuals whose ECGs have all waveforms flattened. Here, we see that the range of values for both FPC1 and FPC2 scores are generally lower than the reference group. Such an approach has been described in \cite{Minhas2025}, who proposed a functional gait deviation index, where multivariate FPCA on a reference dataset is obtained, and projected scores for new subjects are scaled and combined to create a subject-level index. \\

We now describe how projection scores are obtained. In FPCA, where each subject provides a single curve, the estimation of projected scores is straightforward and can be obtained in the \texttt{refund} package in \texttt{R} \citep{refundR}. Given a new functional observation $y_i^{\text{new}}(t)$, the function is first centred using the mean function $\mu(t)$ estimated from the reference dataset, and then projected onto the $K_1$ eigenfunctions $\phi_{k_1}$, for $k_1 \in \left\{1, 2, ..., K_1 \right\}$, obtained from the FPCA of the reference dataset. The estimated FPCA scores are computed as \citep{Yao2005}:

\begin{equation}
\label{fpca_projection_score}
\xi_{k_1}^\text{new} = \int \left\{ y_i^{\text{new}}(t) - \mu(t) \right\} \,\phi_{k_1}(t)\,\mathrm{d}t.
\end{equation}

\noindent Obtaining projection scores in MFPCA is more complex, as MFPCA decomposes variability across two (or more) hierarchical levels, each with its own eigenfunctions. Furthermore, there is currently no existing function to compute them in FDA-related \texttt{R} packages. For MFPCA, projection requires assigning the functions from the new dataset to the appropriate levels of the hierarchy (e.g. between- or within-subject), centring functions at each level using the corresponding reference means, and estimating or conditioning on random-effect scores at higher levels before computing lower-level scores.\\

We describe the procedure for estimating multilevel functional principal component scores from a new set of curves recorded on a shared time grid \citep{Di2009}. Since observations are measured on a common grid, we approximate inner products $\langle \cdot, \cdot \rangle$ using equal weights
$w_\ell = 1/L$, where $L$ is the number of timepoints:
\begin{equation*}
\langle f, g \rangle =
\int_T f(t) g(t) dt
\approx
\sum_{\ell=1}^L f(t_\ell) g(t_\ell) w_\ell .
\end{equation*}

Given functional data $y_{ij}^{new}(t)$ from new a subject $i$ and curve $j$, we define the raw level-1 projection
\begin{equation*}
c_{ijk_{1}}
=
\langle y_{ij}^{new} - \mu, \phi^{(1)}_{k_1} \rangle
=
\sum_{\ell=1}^L
\big(y_{ij}^{new}(t_\ell) - \mu(t_\ell)\big)
\phi^{(1)}_{k_1}(t_\ell) w_\ell .
\end{equation*}

\noindent For a new subject $i$,  for $i \in \left\{ 1, 2, \cdots N \right\}$ with $n_i$ waves, the subject-average projection is
\begin{equation*}
\bar c_{ik_1}
=
\frac{1}{n_i}
\sum_{j=1}^{n_i} c_{ijk_1}.
\end{equation*}

\noindent Under the MFPCA model, this quantity decomposes into the true between-subject score $b_{ik_1}$ and an averaged residual error term $\bar\varepsilon_{ik_1}$: 

\begin{align*}
\bar c_{ik_1} &= b_{ik_1} + \bar\varepsilon_{ik_1},\\
\bar\varepsilon_{ik_1} &\sim N\!\left(0,\frac{\sigma_e}{n_i}\right).
\end{align*}

The empirical best linear unbiased predictor (EBLUP) \citep{Henderson1975} of the
between-subject scores $b_{ik_1}$ is
\begin{equation*}
\widehat b_{ik_1}=
\frac{\hat{\lambda}^{(1)}_{k_1}}
{\hat{\lambda}^{(1)}_{k_1} + \sigma_e / n_i} \bar c_{ik_1}.
\end{equation*}

\noindent which subject-level projections are shrunk toward zero, with the degree of shrinkage determined by the relative magnitudes of the eigenvalue $\lambda^{(1)}_{k_1}$, the residual variance $\sigma_e$, and the number of curves $n_i$.\\

\noindent The principal component between-subjects fitted for the subject $i$ and the curve $j$ is:
\begin{equation*}
\widehat f^{(1)}_{ij}(t)
=
\sum_{k=1}^{K_1}
\widehat b_{ik_1}\,\hat{\phi}^{(1)}_{k_1}(t).
\end{equation*}
\noindent To obtain within-subject projections, we define the residual functions
\begin{equation*}
r_{ij}(t)
=
y_{ij}^{new}(t) - \mu(t) - \widehat f^{(1)}_{ij}(t),
\end{equation*}
\noindent and raw within-subject projections
\begin{equation*}
\tilde c_{ijk_2}
=
\langle r_{ij}, \hat{\phi}^{(2)}_{k_2} \rangle.
\end{equation*}

Since within-subject scores are curve-specific, we have
\begin{align*}
\tilde c_{ijk_2} &= a_{ijk_2} + \varepsilon_{ijk_2},\\
\varepsilon_{ijk_2} &\sim N(0,\sigma_e).
\end{align*} 
\noindent The EBLUP of the curve-specific score $a_{ijk_2}$ is
\begin{equation*}
\widehat a_{ijk_2}
=
\frac{\hat{\lambda}^{(2)}_{k_2}}
{\hat{\lambda}^{(2)}_{k_2} + \sigma_e}
\;\tilde c_{ijk_2}.
\end{equation*}

\noindent The fitted within-individual principal component for subject $i$ at curve $j$ is then given by
\begin{equation*}
\widehat f^{(2)}_{ij}(t)
=
\sum_{\ell=k_2}^{K_2}
\widehat a_{ijk_2}\, \hat{\phi}^{(2)}_{k_2}(t).
\end{equation*}

\section{Simulation}
\label{Sim}
While pre-defined scalar summaries, such as amplitudes of the T-, R- and P-peaks, may typically be of interest as scalar summaries from digital outcome measures, we compared their performance to data-driven projection scores from an MFPCA, as described in Section \ref{functional}. Motivated by the AW ECG dataset described in Section \ref{ECG}, our simulation study aimed to compare these two approaches to generating scalar summaries from ECG data in terms of: 

\begin{enumerate}
    \item known-groups discrimination, specifically if they can discriminate between a healthy and diseased group; 
    \item reliability (intra-cluster correlation coefficient).
\end{enumerate}

\subsection{Data Generating Mechanism}

To ensure our simulated data closely mirrored real-world clinical scenarios, we grounded our simulation in the ECG dataset that is our running example. As illustrated in Figure \ref{initial_fpca}, we first performed a two-level MFPCA on AW ECG recordings from 59 healthy participants. We then used the estimated components from this model to generate realistic simulated ECG data. We denote by $\hat{\mu}(t)$ the estimated mean function, while $\hat{\phi}_{k_1}^{(1)}(t)$  and $ \hat{\phi}_{k_2}^{(2)}(t)$ denote the estimated between-subject and within-subject eigenfunctions, respectively. \\

We denote by $\hat{\eta}_{ik_1}$ and $\hat{\xi}_{ijk_2}$ the empirical between-subject and within-subject scores, respectively, for $i \in \left\{1, .., 59 \right\}, k_1 \in \left \{1, ..., K_1 \right\}, j \in \left\{ 1, ..., n_i \right\}$ and $k_2 \in \left \{1, ..., K_2 \right\}$. The sample mean vector and covariance matrix of the empirical within-subject scores are given by: 
\begin{equation*}
\bar{\xi} = \mathrm{E}[\hat{\xi}_{ij}], 
\qquad 
\Sigma_{\xi} = \mathrm{Cov}(\hat{\xi}_{ij}).
\end{equation*}

\noindent To simulate ECG data from a healthy population, we first obtain a bootstrap resample of between-subject scores, which we denote $\eta_{ik_1}^{\ast}$. The bootstrap resampling method ensures that the relationship between the scores and the estimated eigenfunctions remains consistent with the original data. As eigenfunctions are uniquely defined only up to a sign change, drawing scores from a multivariate normal distribution at the between-subject level can lead to inverted peaks, which are uncharacteristic of a healthy population. \\

Then, for each subject, to generate ECG curves from two distinct measurement periods (i.e. Day 1 and Day 2), within-subject score vectors are drawn from the multivariate normal distribution, for $m \in \left\{1, 2 \right\}:$

\begin{equation*}
\xi_{ij}^{\ast(m)} \sim \mathcal{N}(\bar{\xi},\, \Sigma_{\xi}).
\end{equation*}

This induces shared subject-level structure but independent day-to-day variability. Finally, the simulated ECG trajectory for subject $i$ and wave $j$ on measurement occasion $m$ were constructed as

\begin{equation}
\label{healthy_simulated}
y_{ij}^{\ast(m)}(t)
= \hat{\mu}(t)
+ \sum_{k_1=1}^{K_1} \eta_{ik_1}^{\ast(m)}\, \hat{\phi}_{k_1}^{(1)}(t)
+ \sum_{k_2=1}^{K_2} \xi_{ijk_2}^{\ast(m)}\, \hat{\phi}_{k_2}^{(2)}(t),
\end{equation}

\noindent where $i \in \left\{1, 2, ..., 59\right\}, j \in \left\{ 1, ..., n_i \right\}$ and $ m \in \left\{ 1,2 \right\}$. \\

We simulated a dataset with two groups of individuals with 59 individuals in each group (the number of individuals in each group matched that of the exemplar dataset). ECG data from the first group, assumed to be from a healthy population, were simulated as in Equation \eqref{healthy_simulated}. ECG data from the second group were initially generated with Equation \eqref{healthy_simulated}, but with one specific change induced. The changes include:  

\begin{itemize}
    \item No change, i.e. healthy individuals; 
    \item Flattened T-waves, which occurs in cardiac conditions such as Hypokalemia \citep{Wang2018};
    \item Changing T-wave amplitudes which can occur in conditions such as Hypokalemia  \citep{Wang2018}; we simulate a simple setting where the extent of flattening can take two values;
    \item Flattening of all waves, which can occur in Amyloidosis \citep{Nijjer2010};
    \item Elevation of the ST-segment, which can be indicative of ST-Elevation Myocardial Infarction \citep{Thygesen2018}.
\end{itemize}

\begin{figure}[]
\centering
\includegraphics[width=\textwidth]{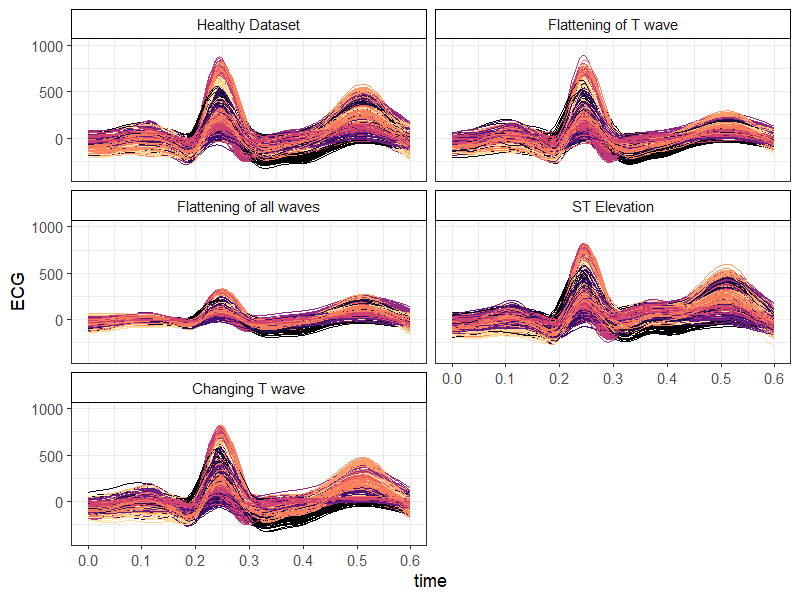}
\caption{Simulated ECG data. Top left: ECGs from a healthy dataset, simulated via the eigenfunctions and eigenvalues obtained from the MFCPA displayed in Figure \ref{initial_fpca}. Top right: ECGs with flattened T-waves induced. Middle left: ECGs with flattening of all waves induced. Middle left: ECGs with ST Elevation induced. Bottom left: ECGs with changing T-wave Amplitudes.}
\label{ECG_simulated_changes}
\end{figure}

These changes are illustrated in Figure \ref{ECG_simulated_changes}. 

To induce change in the second group, one or more Gaussian perturbations were applied on the ECG waveform. A Gaussian perturbation at location $c$ and width $\tau$ is given by
\begin{equation}
    p(t; A, c, \tau) = A \exp\!\left(
    -\frac{1}{2}
    \left( \frac{t - c}{\tau} \right)^{2}
\right).
\end{equation}

\textbf{Flattened T-wave}

To flatten the T-wave, we let $c=0.508$ which corresponds to the location of the T-wave (one of the landmarks) and set $\tau=0.1, A=0.5$ and induce the change as follows:

\begin{equation}
\tilde{y}^{*(m)}_{ij}(t) = y^{*(m)}_{ij}(t) \cdot \left(1-p(t; A=0.5, c=0.508, \tau=0.1) \right). 
\end{equation}

\noindent \textbf{Changing T-wave Amplitudes}

\noindent In this setting, there is heterogeneity in the extent to which the T-wave are flattened. We set $c=0.508, \tau=0.1$ and let $A_i$ either take value 0.2 or 0.7 with equal probability: 

\begin{equation}
\tilde{y}^{*(m)}_{ij}(t) = y^{*(m)}_{ij}(t) \cdot \left(1-p(t; A_i, c=0.508, \tau=0.1) \right). 
\end{equation}

\textbf{All waves flattened}\\
\noindent Here, we induced changes in three locations of the ECG; the P-peak ($c_P=0.104$), the R-peak ($c_R=0.25)$ and T-peak ($c_T=0.508$). The amplitude is set to $A=0.2$ and width is set to $\tau = 0.1$ for all three peaks.
\begin{equation}
\tilde{y}^{*(m)}_{ij}(t) = y^{*(m)}_{ij}(t) \prod_{r \in \left\{P, R, T \right\}} \bigl( 1 - p(t; A, c_r, \tau) \bigr).
\end{equation}

\textbf{ST elevation}\\
\noindent Here, to elevate the ST segment, we let $c=0.37, \tau=0.05, A=120$ and induce the change as follows:

\begin{equation}
\tilde{y}^{*(m)}_{ij}(t) = y^{*(m)}_{ij}(t) \cdot \left(1-p(t; A=120, c=0.37, \tau=0.05) \right). 
\end{equation}

\subsection{Methods}
We considered three methods of summarizing the repeated curves: pre-specified scalars, scores from an FPCA, and scores from an MFPCA. 

\textbf{Scalar approach}\\
Here, we focused on changes in amplitudes in peaks of the ECG. For each subject, we first computed the mean amplitude across repeated curves for the P-, R- and T-waves. Then, the summary metric is the median of the subject-specific mean amplitudes:

\begin{equation}
\bar{a}_i(t)
    = \mbox{median}_i
        \left( 
            \frac{1}{n_i} \sum_{j=1}^{n_i} y_{ij}(t) 
        \right).
\end{equation}

\noindent Since the curves are landmarked at the P-, R- and T-peaks, the P-amplitude corresponds to $\bar{a}(0.104)$, the R-amplitude to $\bar{a}(0.25)$ and T-amplitude to $\bar{a}(0.508)$.

\textbf{Scores from an MFPCA} \\
We estimated projection scores using the MFPCA shown in Figure \ref{initial_fpca} and using the estimation procedure described in Section \ref{Estimation_scores}. We retain the subject-level scores for the first three eigenfunctions, as these explain $91.2\%$ of the total subject-level variability. Figure \ref{plot_projections} provides an illustration of projection scores for the ``Flattening of all waves" and ``Flattening of T wave" scenarios.

\textbf{Scores from a FPCA} \\
We collapse repeated curves within an individual by computing the average curve per individual. Then, using Equation \eqref{fpca_projection_score}, we obtain projection scores relative to the FPCA as shown in Figure 1 in Supplementary File A. We retain scores for the first three eigenfunctions.

\subsection{Performance measures}

For each summary measure, we evaluate known-groups discrimination via:
\begin{enumerate}
    \item the p-value from the Mann-Whitney-U test of differences between two groups;
    \item Area under the Curve (AUC) from a Receiver Operating Characteristics (ROC) analysis. 
\end{enumerate}

To assess test-retest reliability of each summary measure, we estimate ICC(A,1) and ICC(C,1) for summaries obtained at measurement occasion 1 and 2. \\

Simulations were repeated 5000 times and the mean of each performance measure across 5000 simulations were computed. \\

Simulations were performed in \texttt{R} version 4.5.1. 

 \subsection{Results}

 Figure \ref{initial_results} displays results from the simulation study. Results are also provided in Table 1 in Supplementary File A. In the \textit{No Change} scenario, we observe that the P-amplitudes have lower reliability compared to the R- and T-amplitudes, which may be a reflection of inherent variation in physiology as well as between-subject variability in how well P-waves are captured by the AW. This can vary due to differences in electrical axis orientation and conductivity. We observe that the three FPCA scores have slightly reduced reliability compared to the R- and T-amplitudes. The three MFPCA scores have similar reliability to each other and to the R- and T-amplitudes. Furthermore, we observe both with the p-value of the Mann-Whitney-U test and AUC of the ROC that no differences are detected between the two groups, as expected. \\

 In the \textit{Flattened T-wave} scenario, we observe similar patterns to the \textit{No Change} scenario in terms of reliability. In terms of known-groups discrimination, we observe that the T-amplitude, the second and third FPCA scores and the second and third MFPCA scores, are able to detect the change between the two groups. We see from Figure \ref{initial_fpca} Panel (B) that the eigenfunctions corresponding to the two MFPCA scores which discriminate between the two groups reflect changes around the T-wave.\\

 In the \textit{Changing T-wave} scenario, due to the heterogeneity of the T-wave change in the second group, the reliability of summary metrics that are closely related to the T-wave are lower: i.e. T-amplitude, the second and third FPCA scores, and the second and third  MFPCA scores. Further, while the second FPCA and second MFPCA scores are able to discriminate between groups at the $5\%$ level, the third FPCA and third MFPCA scores are no longer able to, demonstrating how reduced reliability can impact known-groups reliability. \\

 In the \textit{All Amplitudes Flattened} scenario, we draw similar conclusions about reliability as in the \textit{No Change} scenario. In terms of known-groups discrimination, we observe that the second and third FPCA scores, as well as the seccond and third MFPCA scores, and the R- and T-amplitudes have similar performance. \\

 Finally, in the \textit{ST Elevation} scenario,  we draw similar conclusions about reliability as in the \textit{No Change} scenario. Here, in terms of known-groups discrimination, none of the pre-specified amplitudes are able to capture the difference between the two groups as these metrics are misspecified for capturing a change such as ST elevation. In contrast, we observe that the third MFPCA score, which we see in Figure \ref{initial_fpca} Panel (B) is capturing change in the region between the R and T-peaks, as well as the third MFPCA score, is able to detect change between the two groups. \\

\begin{figure}[]
\centering
\includegraphics[width=1\textwidth]{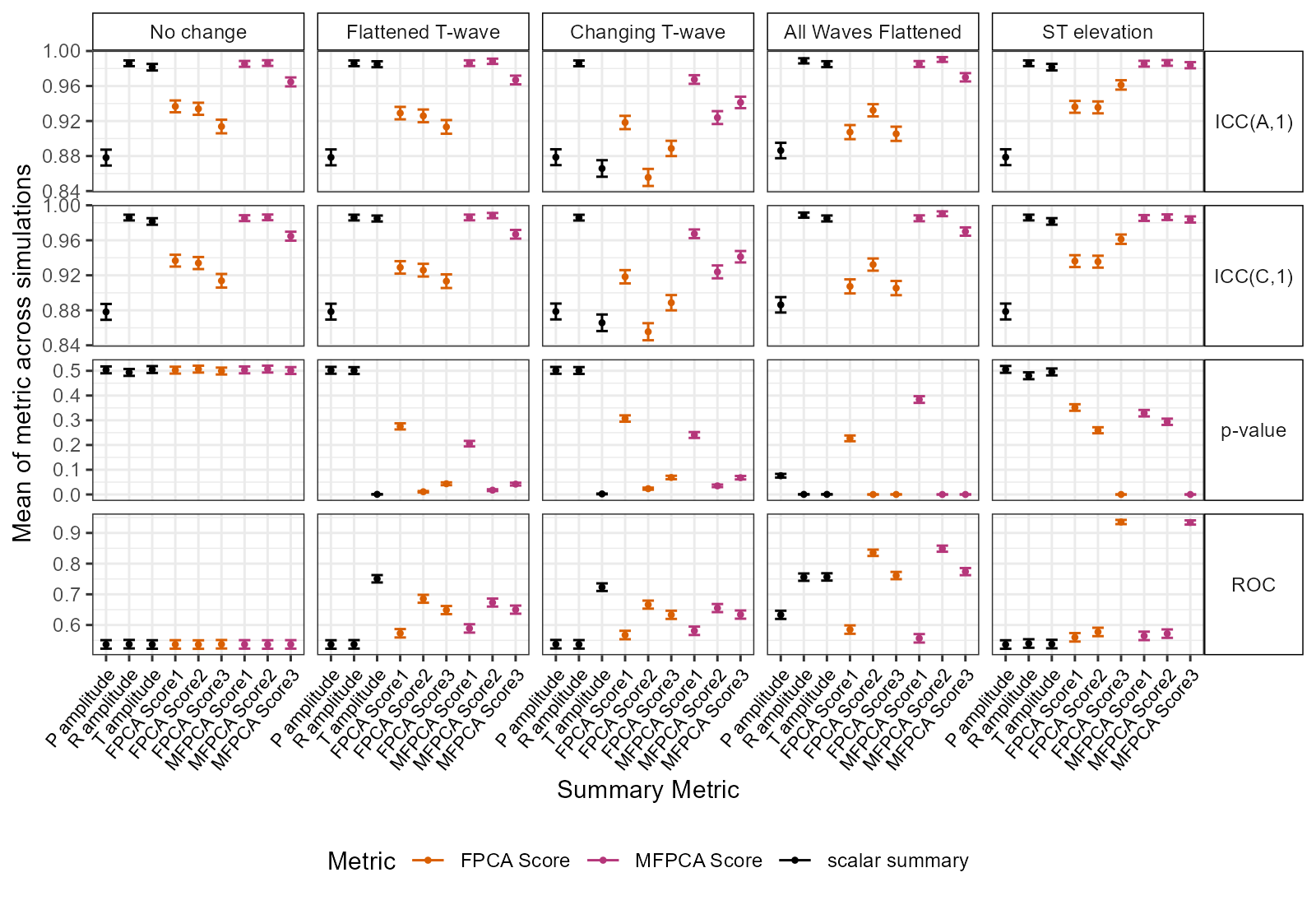}
\caption{Results from the simulation study: across simulation settings (columns) performance in terms of ICC(A,1), ICC(C,1), p-value of Mann-Whitney-U test and ROC is shown (rows) for three types of summary metrics from functional data: pre-specified scalar summaries including the P-, R- and T-amplitudes, scores from an FPCA and scores from an MFPCA.}
\label{initial_results}
\end{figure}

To summarise the key conclusions, we observed that MFPCA scores exhibit higher reliability than FPCA scores derived from averaged curves, because MFPCA explicitly partitions the total variance into between-subject and within-subject components. R- and T-amplitudes were also shown to have good reliability in almost all scenarios, and P-amplitudes were generally less reliable, likely due to greater underlying variability of P-amplitudes. Among all scenarios where a change is induced, at least one of the MFCA scores and one of the FPCA scores were able to discriminate effectively between the two groups, demonstrating the flexibility of the data-driven approaches. Specifically for the \textit{ST elevation} scenario, the amplitude-based scalars were unable to discriminate between groups. Finally, we observed in the \textit{Changing T-wave} scenario that reduced reliability led also to reduced ability to discriminate between groups across scalar summaries and scores.\\

\section{Exemplar 2: Gait Data from People Living with Parkinson's Disease}
\label{PDdata}

\begin{figure}[]
\centering
\includegraphics[width=1\textwidth]{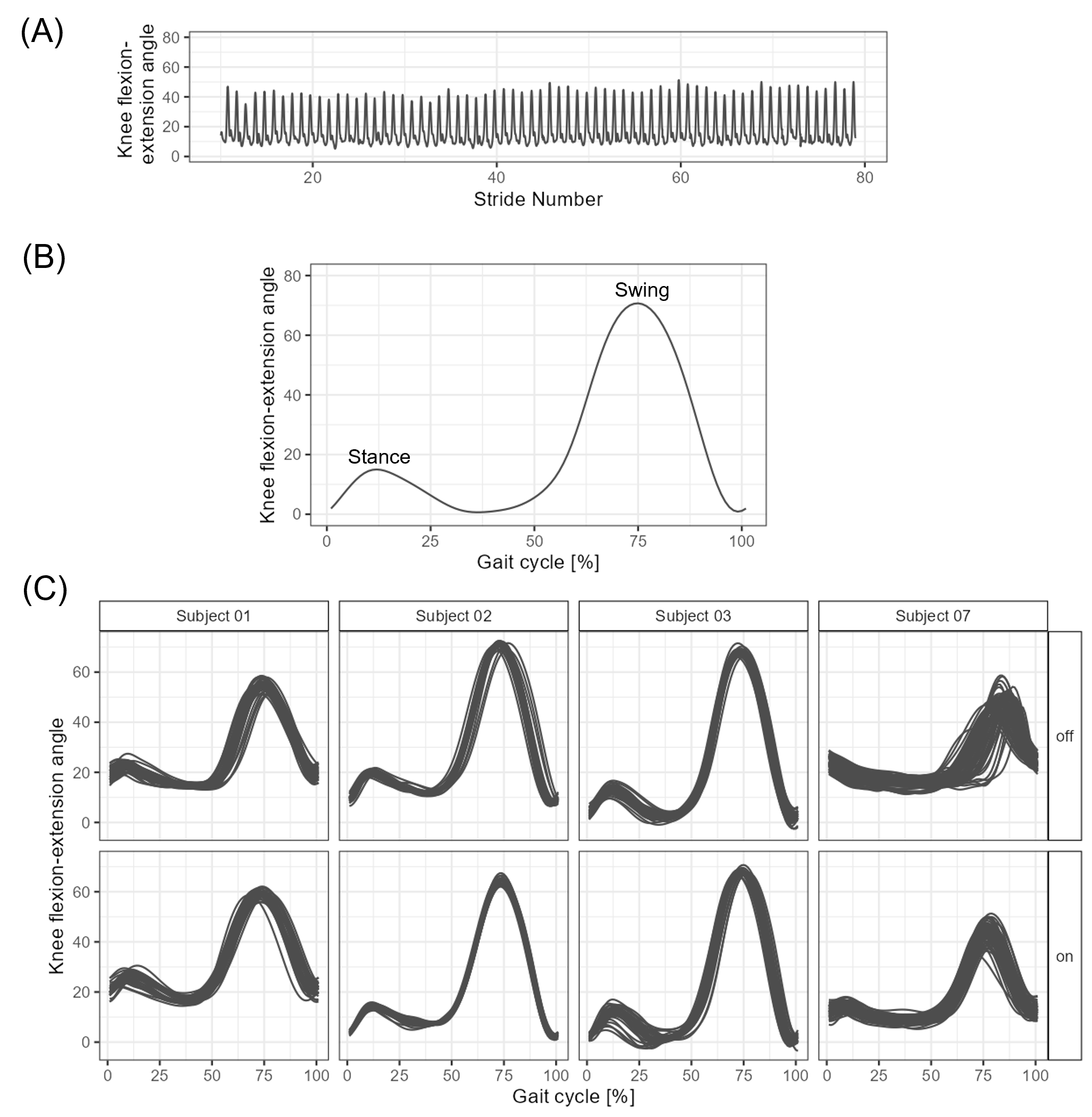}
\caption{Left knee flexion-extension data. (A) A full recording of knee flexion-extension data from one individual. (B) A single curve for one stride with the stance and swing peaks indicated. (C) knee flexion-extension data from four individuals with Parkinson's disease; the angle of knee flexion-extension is plotted against the percentage of gait cycle for each stride. The top panel displays data from a gait analysis when individuals are off medication, and the bottom panel shows measurements when individuals are on medication.}
\label{knee_data}
\end{figure}

We now demonstrate how MFPCA can be used to derive meaningful summary metrics from kinematic data from individuals living with Parkinson's disease (PD). We evaluate the correlation of these scores with gold-standard outcome measures in PD (MDS-UPDRS Part II and III) and their ability to detect change, focusing on the \textit{convergent validity} and \textit{responsiveness to change} aspects of validation. We analyse publicly available data from 22 Parkinson’s Disease (PD) patients, provided by \citep{Shida2023, BoariCoelho2022}, and data from 10 healthy individuals, provided by \citep{Helwig2016_UCI}. The dataset includes multiple kinematic variables, such as pelvic tilt, hip flexion/extension, and knee flexion/extension, measured at 101 equally spaced timepoints throughout the gait cycle from a three-dimensional (3D) motion-capture system. While this does not represent data from a remotely monitored digital health device, the high-frequency data collected has similar characteristics to those obtained by wearable sensors. In the PD dataset, measurements were collected on two separate occasions: once after patients had been off medication for 12 hours, and once while they were on medication. Figure \ref{knee_data} (A), we display left knee flexion-extension angle data for a full recording for one dividual. In (B), we display a single curve for one stride; we observe two peaks related to the stance and swing phases of gait. In (C), we observe multiple curves from four individuals, each on two occasions: in the ``off" medication state and the ``on" medication state. We observe variability between individuals, for example in the timing and shape of the peak, as well as variability within individuals; there is greater variability in the shape of the curves in the ``off'' medication state for subjects 02 and 07 compared to their ``on'' medication state. The PD dataset has previously been examined using a multivariate (but not multilevel) functional approach \citep{Minhas2025}.\\

In the PD dataset, in addition to kinematic measurements, the gold-standard PD outcomes were measured at the on- and off-medication occasions. Specifically, the Movement Disorder Society’s Unified Parkinson’s Disease Rating Scale (MDS-UPDRS) Part II, which evaluates motor experiences during daily activities, and Part III, which assesses motor functions including rigidity and agility, were recorded. Higher scores on these scales reflect greater impairment.\\

\begin{figure}[]
\centering
\includegraphics[width=1\textwidth]{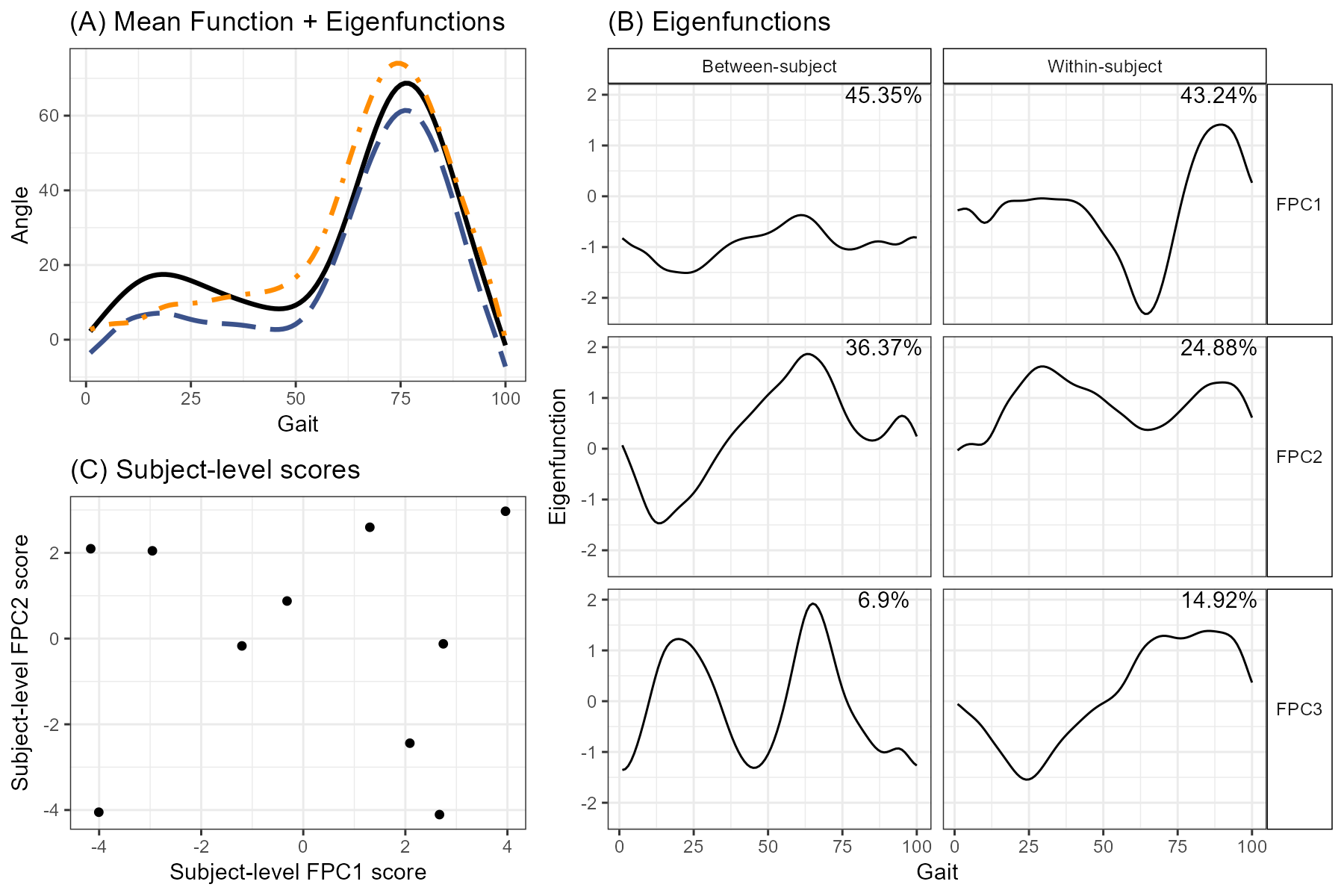}
\caption{Results from MFPCA on 10 healthy individuals' data on knee flexion/extension. We display the overall mean function (top) and the estimated between- and within-subject eigenfunctions (bottom).  }
\label{PD_healthy_fpca}
\end{figure}

To begin with, we display results of an MFPCA on the 10 healthy individuals' knee flexion/extension data in Figure \ref{PD_healthy_fpca}. This serves as the reference dataset  and the MFPCA from which we obtain projections for the PD data. We observe that the first between-subject eigenfunction shows fluctuations around zero, possibly reflecting shifts in amplitude across individuals. The second and third between-subject eigenfunctions have large peaks and troughs, likely indicating variations in the timing and shape of the two peaks in the mean function. The first three eigenfunctions explain 88.6\% of the total variability at the between-subjects level. 

\subsection{Discriminating Between Individuals at Different Stages of PD}
\begin{figure}[]
\centering
\includegraphics[width=0.85\textwidth]{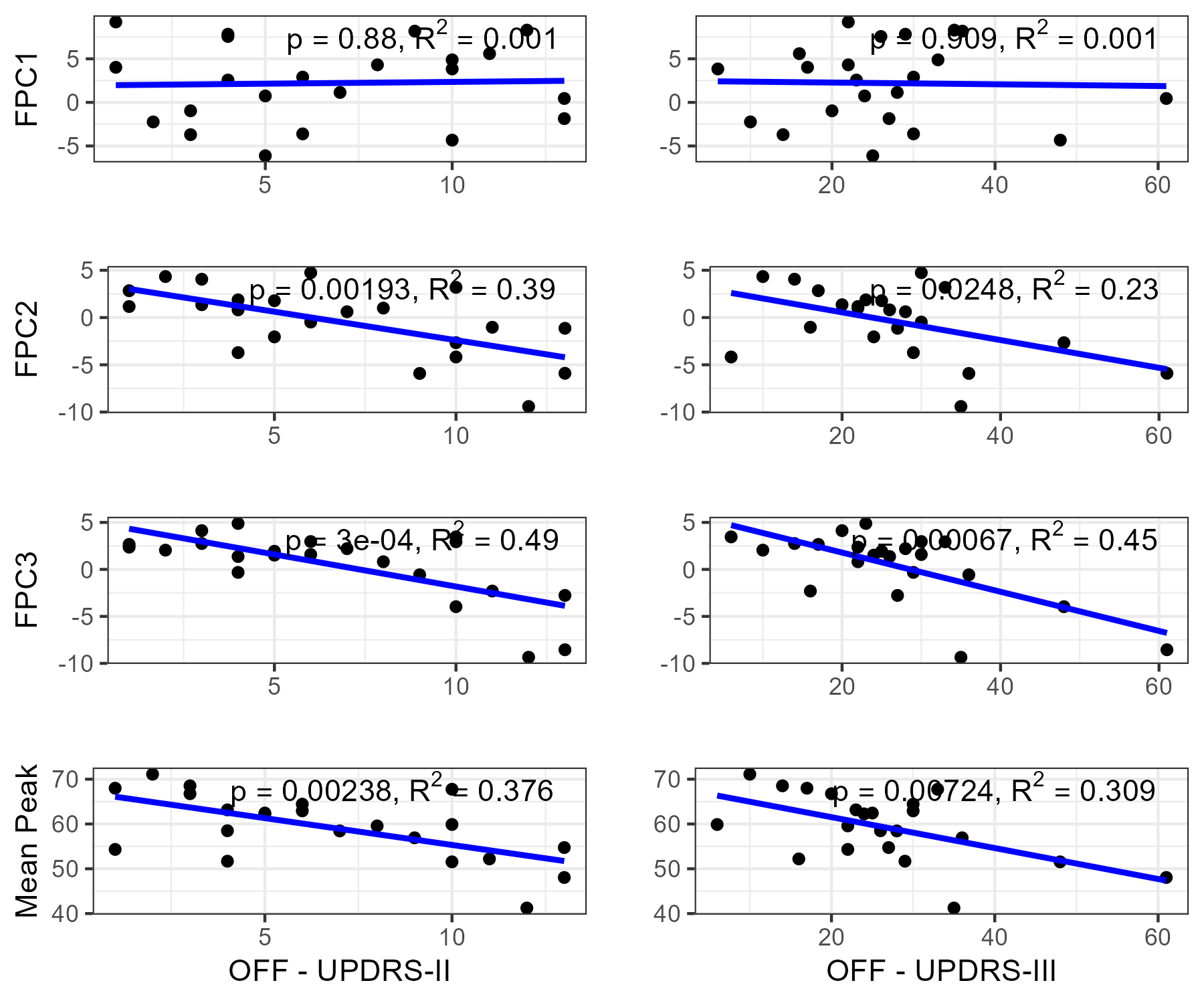}
\caption{Assessment of convergent validity of of different summaries of knee flexion-extension data. Scores from the first three principal components of an MFPCA, as well as the mean of the swing peak, are regressed against the gold-standard MDS UPDRS Part II (on the left) and MPS UPDRS Part III (on the right).}
\label{PD_construct_validity_single}
\end{figure}

For the 22 PD patients, we analyse the knee flexion-extension data when they are off medication, and compute their between-subject scores by projecting the curves onto the MFPCA derived from the healthy dataset. In Figure \ref{PD_construct_validity_single}, we regress the between-subject scores from the first three principal components against the MDS-UPDRS Part II score (left) and MDS-UPDRS Part III score (right). Furthermore, we include the mean amplitude of the peak of the swing phase of the gait cycle (from here on referred to as ``mean peak"), as a simple summary metric, to compare against MFPCA scores. We report the $R^2$ coefficient and the corresponding p-value for the regression slope. We observe that the scores from the second and third eigenfunctions, as well as the mean peak, correlate strongly with gold-standard outcome measures.

\subsection{Assessing Sensitivity to Change}
\begin{figure}[]
\centering
\includegraphics[width=0.85\textwidth]{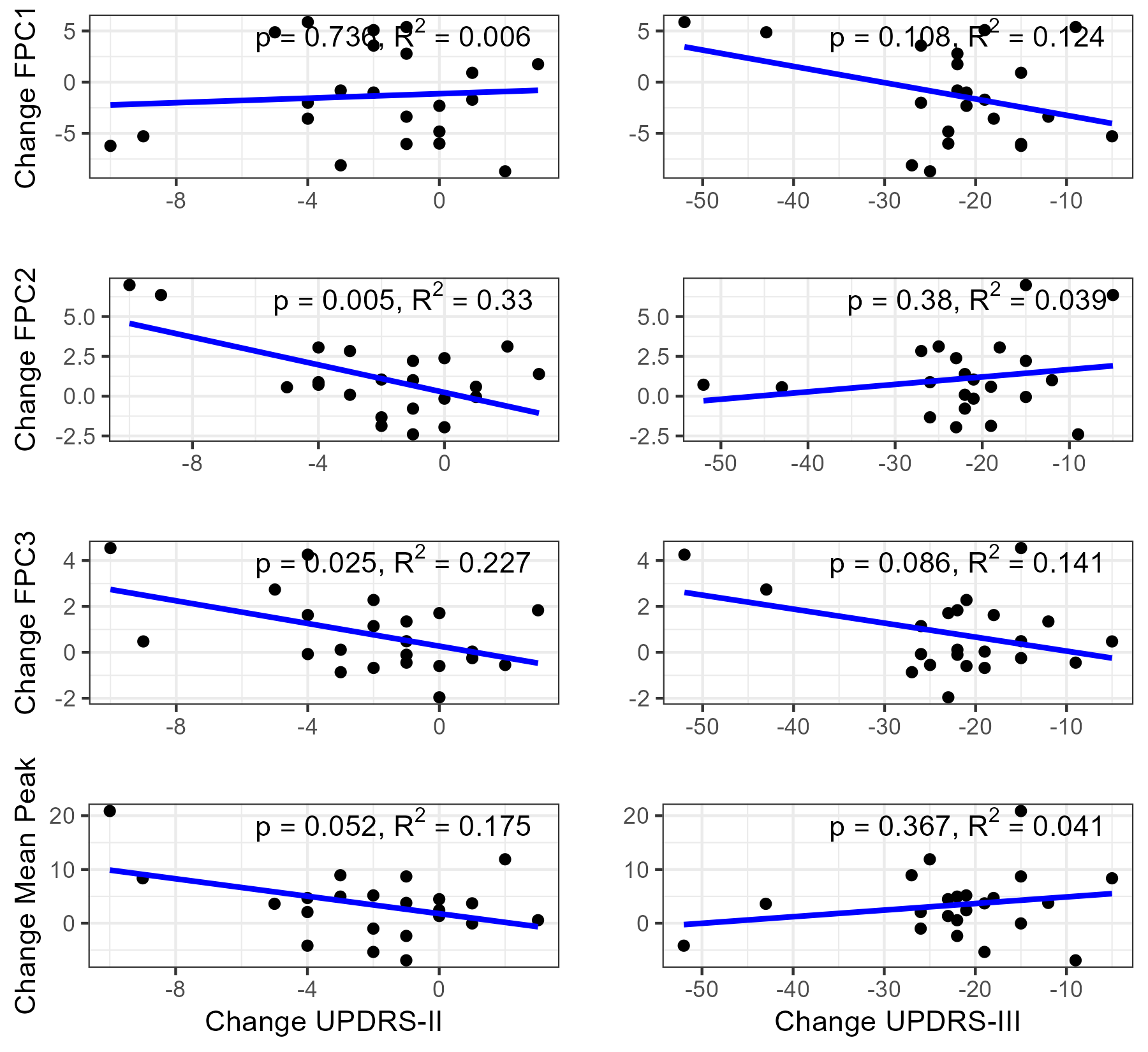}
\caption{Assessment of responsiveness to change of different summaries of knee flexion-extension data. Change scores from the first three principal components of an MFPCA, as well as the change in mean of the swing peak, are regressed against the change in gold-standard MDS UPDRS Part II (on the left) and MPS UPDRS Part III (on the right).}
\label{PD_change_scores}
\end{figure}

In the previous section, we obtained between-subject scores when individuals are off medication by obtain projection scores relative to an MFPCA derived from a healthy dataset. We can repeat this procedure with data when individuals are on medication, obtaining two projection scores for each individual for their gait on and off medication. Then, we compute change scores by taking the difference between the MFPCA projection scores on and off medication. Analogously, we compute the changes in the gold-standard outcomes, MDS-UPDRS Part II, MDS-UPDRS Part III, as well as change in mean peak. In Figure \ref{PD_change_scores}, we regress the MFPCA change scores against the change in MDS-UPDRS Part II (left panel) and change in MDS-UPDRS Part (III) (right panel). We report the $R^2$ coefficient and the corresponding p-value for the regression slope. We observe that changes in the second and third MFPCA scores are significantly correlated with the change in MDS-UPDRS Part II at the $5\%$ level, while the change in mean peak is not significant at the $5\%$ level. None of the summary metrics are significantly correlated with MDS-UPDRS Part III.

Overall, we observe that both the second and third MFPCA scores correlate well with both of the MDS-UPDRS scores; particularly for MDS-UPDRS Part II, and these two functional scores lead to increased $R^2$ compared to the mean peak. Further, when considering change due to medication, the change in the second MFPCA score had improved correlation compared to the change in mean peak, demonstrating the potential utility of MFPCA scores.

\section{Discussion}
\label{Discussion}
Digital outcome measures provide a vast quantity of physiological data which is often functional and hierarchical in nature. While data-driven approaches to obtaining outcome measures are generally less common in clinical trials, interest in their use, particularly as additional or supportive analyses, is increasing \citep{Zablocki2024, Sanchez2014, Tackney2024}. To obtain data-driven summary metrics which are comparable across independent trials, we took the approach of normative modelling, where data from a reference population was used to derive a standardised functional subspace; participants' data were projected onto this reference space to obtain scores that quantify their deviation from the healthy functional profile. These projection scores quantify how each individual deviates from the reference population in principal directions of variation. Our investigation demonstrated that MCFPA scores are a viable approach to obtaining efficient summaries from functional data. Regardless of the functional form of the change, including local and global changes, at least one of the MCFPA scores effectively distinguished the treatment group from the control.\\ 

Specifically, our investigation compared the performance of MFPCA and FPCA scores against pre-specified scalars in two settings: firstly in a simulation study based on ECG data to assess reliability and ability to discriminate between groups, and secondly in an analysis of gait-related data in PD patients to assess convergent validity and responsiveness to change. In the simulation study, we demonstrated that MFPCA scores were more reliable than FPCA scores obtained after averaging repeated curves from individuals. Amongst the first three MFCPA scores, at least one was able to detect change between the two groups in all scenarios were changes were induced. Pre-specified summaries (P-, R- and T-peaks) performed well in terms of reliability and ability to detect change when they were well specified for the induced change, but were unable to detect change when they were misspecified. In the analysis of gait-related data, we demonstrated that at least one of the MFPCA scores were well correlated to the MDS-UPDRS scores II and III at a specific time point. Furthermore, when there was change due to medication, changes in one of the MFPCA scores was well correlated with  changes in the MDS-UPDRS part II score. While we constructed MFPCA scores assuming a specific configuration of a functional multilevel model, other random effect configurations and structural assumptions may be considered \citep{Gaynanova2022}.\\

We note a number of limitations of our investigation. Firstly, the reference datasets have small sample sizes with 59 healthy individuals in the ECG setting and 10 healthy individuals in the gait example. Providing recommendations on the size of a suitable reference set is an area of future work. Increasingly, the availability of large-scale population studies with data from wearable technologies, such as the UK Biobank, as well as emerging validation datasets on digital outcome measures, offer a unique opportunity to establish normative reference datasets. Secondly, our analyses were restricted to the first three FPCA and MFCPA scores, as they explained 91.2\% and 88.6\% of the total variation, respectively. Future work could include sensitivity analyses to the number of principal component scores included. Thirdly, uncertainty around the scores were not addressed in our work. This could be accounted for via a fully Bayesian semiparametric approach \citep{Chen2026}.\\

MFPCA provides distinct methodological advantages for the different clinical trial phases. As exploratory outcomes in early phase trials, the potential for MFPCA scores to obtain more sensitive summary metrics is a strategic advantage. In early phase trials, patient cohorts are often small and traditional scalar metrics may lack power to detect a treatment effect. Therefore, the potential for MFCPA scores to detect morphological changes more sensitively is a promising direction. This was illustrated in the PD example where changes in FPC2 and FPC3 had higher correlation to changes in the MDS UPDRS Part II score than the mean peak. Furthermore, MFPCA offers a useful framework for simulating digital outcomes on the granular level, as demonstrated by the data generating mechanism of our simulation study, which helps support evaluation of study designs.\\

If the goal is to establish a validated digital outcome measure for a pivotal Phase III trial as a clinical outcome assessment, clinical interpretability becomes more crucial due to regulatory considerations. Here, future work in translating mechanistic insights from the eigenfunctions to simpler, interpretable functions, and corresponding scalar summaries, may be needed. Furthermore, in late-phase trials, a critical step for demonstrating validity of digital outcomes is defining the Minimal Clinically Important Difference (MCID). Here, changes in principal component scores (or clinically interpretable summary metrics derived from them) can be anchored to improvements in patient health status, such as via a Global Rating of Change (GRC) scale \citep{Mccarthy2023}. A ROC curve analysis to identify the specific cut-point on the principal component axes which discriminates between patients who report subjective clinical improvement and those who do not, could be performed to define the MCID for functional scores. \\

Finally, there are specific areas of future work for different types of functional outcomes. In the specific setting of ECG curves, we took the approach of landmarking cardiac cycles at the P, R, and T waves, which removed the temporal positioning of these peaks. This variability in phase may itself be clinically informative. Future work should explore methods which can capture both phase and amplitude variation. Furthermore, the examples in this article were restricted to the analysis of curves over a temporal domain. For digital health outcomes  such as glucose readings from Continuous Glucose Monitoring, where fluctuations are driven by sporadic events like meals or exercise rather than a rhythmic signals such as ECG or gait, distributional representations via functional data analysis is a promising approach for defining digital biomarkers \citep{Matabuenaetal2026}.\\

\section*{Funding}
MST, Advanced Fellow, NIHR305417, is funded by the National Institute of Health and Care Research for this research project. The views expressed are those of the authors and not necessarily those of the NIHR or the Department of Health and Social Care. MP is supported by the Ulverscroft Vision Research Group (UCL). JC and SSV are supported by MRC in partnership with NIHR (CCTI grant number UKRI934). JM is supported by the Else Kröner-Fresenius-Stiftung (project number: 2023\_EKES.04), a grant of the Deutsche Herzstiftung e.V. (German Heart Foundation), an unrestricted research grant by the Bayer AG, and research grants of the DFG (MU 4555/2-1, project number: 455425596; and MU 4555/5-1, project number: 546575044). TK is supported by the ReForM program of the University of Regensburg.

\section*{Supplementary material}
A GitHub repository contains \texttt{R} code to reproduce simulations: \url{https://github.com/mst1g15/MFPCA_digital_outcomes}. A vignette which illustrates the analysis of gait data from individuals living with Parkinson's Disease is provided: \url{https://mst1g15.github.io/MFPCA_digital_outcomes/}. 

\section*{Conflicts of interest}
SSV is on the advisory board for PhaseV (unrelated to this work). JM reports compensation received for public talks related to cardiac amyloidosis and/or hypertrophic cardiomyopathy by Bayer, Bristol Myers Squibb, Eli Lilly, Alnylam, Boehringer Ingelheim, and Astra Zeneca. JM also reports an unrestricted scientific grant by the Bayer AG. JM is a cofounder of Corgene GmbH, which is devoted to cardiac gene therapies. TK reports compensation received for public talks related to cardiac amyloidosis and/or hypertrophic cardiomyopathy by Bayer, Pfizer, Alnylam and Astra Zeneca. TK also recieved travel funding for scientific congresses by Bayer, Eli Lilly and Pfizer.

\section*{Appendix}

\subsection{Description of Data Pre-processing}
A 30 second ECG was recorded with an Apple Watch (Series 9). The raw ECG data was exported and pre-processed prior to analysis. On some occasions, ECG recordings appeared inverted due to participants wearing the watch on the opposite wrist. To address this, inverted ECGs were corrected via the neurokit2 \texttt{ecg\_invert()} command \citep{Makowski2021neurokit}. \\

We used a modified version of a template-matching algorithm \citep{Orphanidou2015} to perform quality control of ECG curves. The original template-matching algorithm classifies the entire ECG recording as good or poor quality. Since ECG recordings typically had poor quality in specific regions, we modified the algorithm to enable removal of individual poor-quality cardiac cycles. We excluded cardiac cycles if their correlation coefficient compared to the template QRS was less than 0.9. We retained good quality complexes if the ECG recording contained 10 or more good quality complexes, otherwise, the ECG recording was deemed poor quality.\\

Following the pre-processing steps, the timing and amplitudes of the P-, R- and T-peaks and onsets/offsets identified by \texttt{ecg\_process()} for each cardiac cycle.\\

Functional outliers were identified and removed using the Functional Boxplot with Modified Band Depth (MBD) method, via the \texttt{fbplot{}} function in the \texttt{fda R} package \citep{fdaR}. 

The remaining cardiac cycles were converted into smooth  functions using a B-spline basis expansion with 18 functions. We performed landmark registration of the ECG curves using the P, R and T-peaks as landmarks. Any ECG curve where all three peaks could not be detected, or were detected in implausible locations along the time domain, were discarded. Using the \texttt{landmarkreg()} from the \texttt{fda R} package, the time axis of each cardiac curves was non-linearly warped so that these three peaks were aligned in time. A roughness penalty ($\lambda = 10^{-5}$) was applied to the warping function to ensure that the lower-amplitude features, such as the P-wave, were aligned.\\

\subsection{}
\begin{figure}[H]
\centering
\includegraphics[width=1\textwidth]{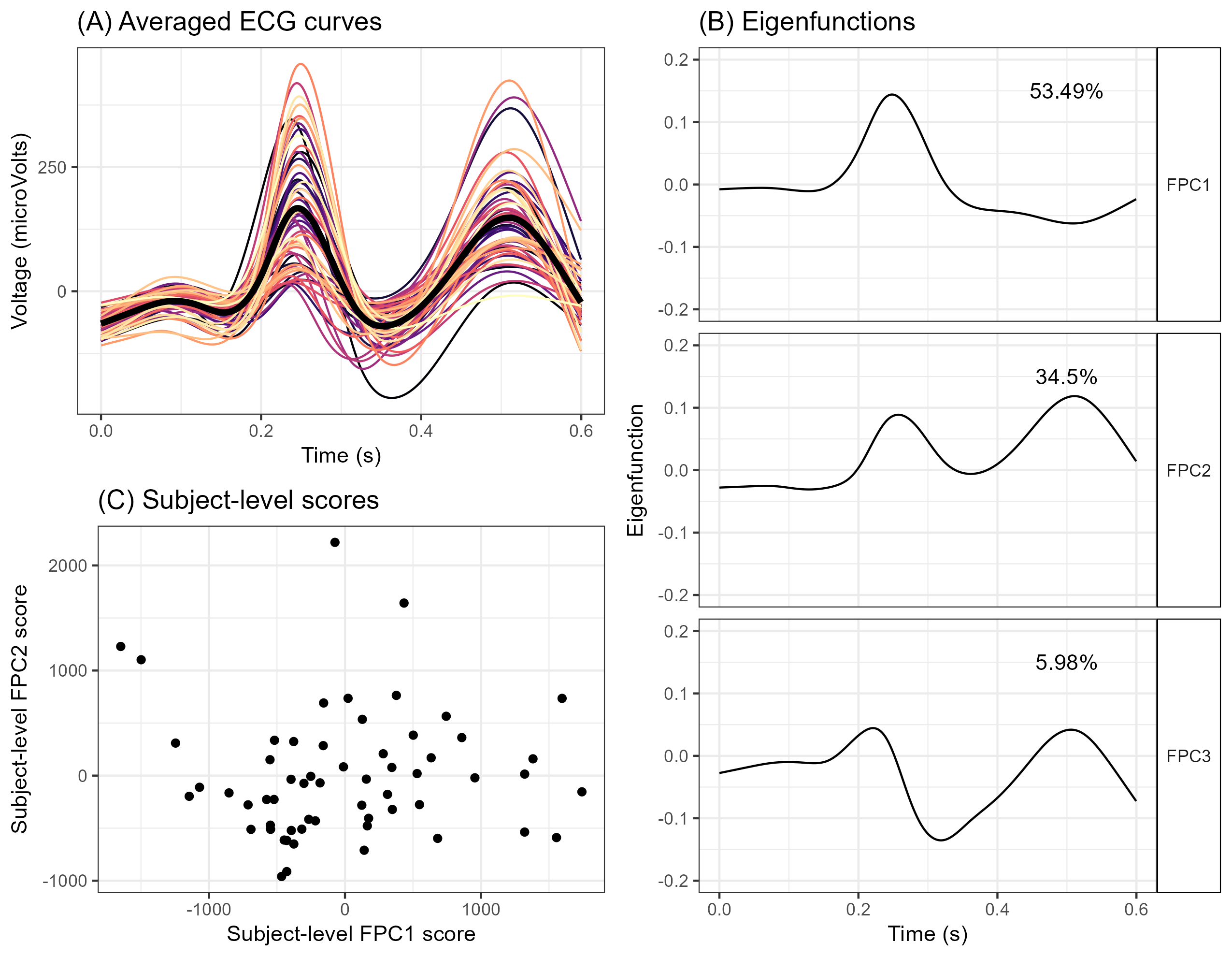}
\caption{Results from a FPCA performed on Apple Watch ECGs from 59 healthy participants, where an average curve is computed for each participant. (A) Landmarked averaged ECGs for each individual with colors distinguishing individuals; the overall mean function is superimposed in black. (B) The first four eigenfunctions. (C) The proportion of variance explained by each FPC. (D) Subject-level FPC2 scores plotted against FPC1 scores.}
\label{initial_single_fpca}
\end{figure}

\bibliographystyle{plainnat}  
\bibliography{library} 

@article{Goldsmith2015,
  title={Generalized multilevel function-on-scalar regression and principal component analysis},
  author={Goldsmith, Jeff and Zipunnikov, Vadim and Schrack, Jennifer},
  journal={Biometrics},
  volume={71},
  number={2},
  pages={344--353},
  year={2015},
  publisher={Wiley},
  doi={10.1111/biom.12278},
  url={https://doi.org/10.1111/biom.12278}
}

@misc{Zhou2024,
  title={Analysis of Active/Inactive Patterns in the NHANES Data using Generalized Multilevel Functional Principal Component Analysis},
  author={Xinkai Zhou and Julia Wrobel and Ciprian M. Crainiceanu and Andrew Leroux},
  year={2024},
  eprint={2311.14054},
  archivePrefix={arXiv},
  primaryClass={stat.ME},
  url={https://arxiv.org/abs/2311.14054}
}

@book{RamsaySilverman2005,
  author    = {Ramsay, J. O. and Silverman, B. W.},
  title     = {Functional Data Analysis},
  edition   = {2},
  publisher = {Springer},
  address   = {New York},
  year      = {2005}
}

@article{Matabuena2023,
  author = {Matabuena, Marcos and Karas, Marta and Riazati, Sherveen and Caplan, Nick and Hayes, Philip R.},
  title = {Estimating Knee Movement Patterns of Recreational Runners Across Training Sessions Using Multilevel Functional Regression Models},
  journal = {The American Statistician},
  year = {2022},
  volume = {77},
  number = {2},
  pages = {169--181},
  doi = {10.1080/00031305.2022.2105950},
  url = {https://doi.org/10.1080/00031305.2022.2105950}
}

@article{Yao2005,
  author = {Fang Yao and Hans-Georg M{\"u}ller and Jane-Ling Wang},
  title = {Functional Data Analysis for Sparse Longitudinal Data},
  journal = {Journal of the American Statistical Association},
  volume = {100},
  number = {470},
  pages = {577-590},
  year = {2005},
  publisher = {Taylor \& Francis},
  doi = {10.1198/016214504000001745},
  URL = {https://doi.org/10.1198/016214504000001745},
  eprint = {https://doi.org/10.1198/016214504000001745}
}

@article{Matabuenaetal2026,
  author    = {Matabuena, Marcos and Crainiceanu, Ciprian M.},
  title     = {Multilevel functional distributional models with applications to continuous glucose monitoring in diabetes clinical trials},
  journal   = {The Annals of Applied Statistics},
  year      = {2026},
  volume    = {20},
  number    = {1},
  pages     = {476--495},
  doi       = {10.1214/26-AOAS2139}
}

@article{Matabuena2026,
  author    = {Matabuena, M. and Sartini, J. and Gude, F.},
  title     = {Beyond scalar metrics: functional data analysis of postprandial continuous glucose monitoring in the AEGIS study},
  journal   = {BMC Medical Research Methodology},
  year      = {2026},
  volume    = {26},
  number    = {1},
  pages     = {39},
  doi       = {10.1186/s12874-025-02748-2},
  url       = {https://doi.org/10.1186/s12874-025-02748-2}
}

@article{Di2009,
  author = {Di, Chong-Zhi and Crainiceanu, Ciprian M. and Caffo, Brian S. and Punjabi, Naresh M.},
  title = {Multilevel functional principal component analysis},
  journal = {Annals of Applied Statistics},
  volume = {3},
  number = {1},
  pages = {458--488},
  year = {2009},
  month = {March},
  doi = {10.1214/08-AOAS206},
  url = {https://doi.org/10.1214/08-AOAS206}
}

@article{Loeve1945,
  author  = {Lo{\`e}ve, Michel},
  title   = {Fonctions al{\'e}atoires de second ordre},
  journal = {Revue Scientifique},
  volume  = {83},
  pages   = {297--303},
  year    = {1945}
}

@article{Karhunen1947,
  author  = {Karhunen, Kari},
  title   = {Über lineare Methoden in der Wahrscheinlichkeitsrechnung},
  journal = {Annales Academiae Scientiarum Fennicae. Series A. I. Mathematica},
  volume  = {37},
  pages   = {1--79},
  year    = {1947}
}

@article{Di2014,
  author       = {Chongzhi Di and Ciprian M. Crainiceanu and Wolfgang S. Jank},
  title        = {Multilevel sparse functional principal component analysis},
  journal      = {Stat},
  volume       = {3},
  number       = {1},
  pages        = {126--143},
  year         = {2014},
  doi          = {10.1002/sta4.50},
  url          = {https://doi.org/10.1002/sta4.50}
}

@article{Orphanidou2015,
  author    = {Christina Orphanidou and Peter J. Charlton and Duncan A. Clifton and Lionel Tarassenko},
  title     = {Signal-Quality Indices for the Electrocardiogram and Photoplethysmogram: Derivation and Applications to Wireless Monitoring},
  journal   = {IEEE Journal of Biomedical and Health Informatics},
  volume    = {19},
  number    = {3},
  pages     = {832--841},
  year      = {2015},
  doi       = {10.1109/JBHI.2014.2320411},
  publisher = {IEEE},
}

@article{Trevethan2017,
   author = {Robert Trevethan},
   doi = {10.1007/s10742-016-0156-6},
   issn = {15729400},
   issue = {2},
   journal = {Health Services and Outcomes Research Methodology},
   month = {6},
   pages = {127-143},
   publisher = {Springer New York LLC},
   title = {Intraclass correlation coefficients: clearing the air, extending some cautions, and making some requests},
   volume = {17},
   year = {2017}
}

@article{Liljequist2019,
   author = {David Liljequist and Britt Elfving and Kirsti Skavberg Roaldsen},
   doi = {10.1371/journal.pone.0219854},
   issn = {19326203},
   issue = {7},
   journal = {PLoS ONE},
   month = {7},
   pmid = {31329615},
   publisher = {Public Library of Science},
   title = {Intraclass correlation – A discussion and demonstration of basic features},
   volume = {14},
   year = {2019}
}

@misc{Ratitch2023,
   author = {Bohdana Ratitch and Andrew Trigg and Madhurima Majumder and Vanja Vlajnic and Nicole Rethemeier and Richard Nkulikiyinka},
   doi = {10.1159/000531054},
   issn = {2504110X},
   issue = {1},
   journal = {Digital Biomarkers},
   month = {8},
   pages = {74-91},
   publisher = {S. Karger AG},
   title = {Clinical Validation of Novel Digital Measures: Statistical Methods for Reliability Evaluation},
   volume = {7},
   year = {2023}
}

@article{Servais2022,
   author = {Laurent Servais and Karl Yen and Maitea Guridi and Jacek Lukawy and David Vissière and Paul Strijbos},
   doi = {10.3233/JND-210743},
   issn = {22143602},
   issue = {2},
   journal = {Journal of Neuromuscular Diseases},
   pages = {335-346},
   pmid = {34958044},
   title = {Stride Velocity 95th Centile: Insights into Gaining Regulatory Qualification of the First Wearable-Derived Digital Endpoint for use in Duchenne Muscular Dystrophy Trials},
   volume = {9},
   year = {2022}
}

@article{Minhas2025,
   author = {Sajal Kaur Minhas and Morgan Sangeux and Julia Polak and Michelle Carey},
   doi = {10.1080/02664763.2025.2514150},
   issn = {13600532},
   journal = {Journal of Applied Statistics},
   publisher = {Taylor and Francis Ltd.},
   title = {The functional gait deviation index},
   year = {2025}
}

@article{Roach2021,
   author = {Koren E. Roach and Valentina Pedoia and Jinhee J. Lee and Tijana Popovic and Thomas M. Link and Sharmila Majumdar and Richard B. Souza},
   doi = {10.1002/jor.24901},
   issn = {1554527X},
   issue = {8},
   journal = {Journal of Orthopaedic Research},
   month = {8},
   pages = {1722-1731},
   pmid = {33615524},
   publisher = {John Wiley and Sons Inc},
   title = {Multivariate functional principal component analysis identifies waveform features of gait biomechanics related to early-to-moderate hip osteoarthritis},
   volume = {39},
   year = {2021}
}

@article{Yoshida2022,
   author = {Kaya Yoshida and Drew Commandeur and Sandra Hundza and Marc Klimstra},
   doi = {10.1016/j.jbiomech.2022.111342},
   issn = {18732380},
   journal = {Journal of Biomechanics},
   month = {11},
   pmid = {36265422},
   publisher = {Elsevier Ltd},
   title = {Detecting differences in gait initiation between older adult fallers and non-fallers through multivariate functional principal component analysis},
   volume = {144},
   year = {2022}
}

@misc{Helwig2016_UCI,
  author       = {Helwig, Nathaniel and Hsiao-Wecksler, Elizabeth},
  title        = {{Multivariate Gait Data}},
  year         = {2016},
  howpublished = {UCI Machine Learning Repository},
  note         = {{DOI}: https://doi.org/10.24432/C5861T}
}

@Manual{fdaR,
    title = {fda: Functional Data Analysis},
    author = {James Ramsay},
    year = {2025},
    note = {R package version 6.3.0},
    url = {https://CRAN.R-project.org/package=fda},
    doi = {10.32614/CRAN.package.fda},
  }

@article{Makowski2021neurokit,
  author  = {Makowski, Dominique and Pham, Tam and Lau, Zen J. and Brammer, Jan C. and Lespinasse, Fran{\c{c}}ois and Pham, Hung and Schölzel, Christopher and Chen, S. H. Annabel},
  title   = {NeuroKit2: A Python toolbox for neurophysiological signal processing},
  journal = {Behavior Research Methods},
  year    = {2021},
  volume  = {53},
  number  = {4},
  pages   = {1689--1696},
  doi     = {10.3758/s13428-020-01516-y}
}

@article{Poleur2021,
  author    = {Poleur, Margaux and Ulinici, Ana and Daron, Aurore and Schneider, Olivier and Dal Farra, Fabian and Demonceau, Marie and Annoussamy, M{\'e}lanie and Vissi{\`e}re, David and Eggenspieler, Damien and Servais, Laurent},
  title     = {Normative data on spontaneous stride velocity, stride length, and walking activity in a non-controlled environment},
  journal   = {Orphanet Journal of Rare Diseases},
  year      = {2021},
  volume     = {16},
  number     = {1},
  pages      = {318},
  doi        = {10.1186/s13023-021-01956-5},
  issn       = {1750-1172},
  url        = {https://doi.org/10.1186/s13023-021-01956-5}
}

@article{Gaynanova2022,
  title={Modeling continuous glucose monitoring (CGM) data during sleep},
  author={Gaynanova, Irina and Punjabi, Naresh and Crainiceanu, Ciprian},
  journal={Biostatistics},
  volume={23},
  number={1},
  pages={223--239},
  year={2022},
  month={01},
  publisher={Oxford University Press},
  doi={10.1093/biostatistics/kxaa023},
  url={https://doi.org/10.1093/biostatistics/kxaa023}
}

@article{Isakadze2020,
title = {How useful is the smartwatch ECG?},
journal = {Trends in Cardiovascular Medicine},
volume = {30},
number = {7},
pages = {442-448},
year = {2020},
issn = {1050-1738},
doi = {https://doi.org/10.1016/j.tcm.2019.10.010},
url = {https://www.sciencedirect.com/science/article/pii/S1050173819301495},
author = {Nino Isakadze and Seth S. Martin}
}

@article{Nazarian2021,
author="Nazarian, Scarlet
and Lam, Kyle
and Darzi, Ara
and Ashrafian, Hutan",
title="Diagnostic Accuracy of Smartwatches for the Detection of Cardiac Arrhythmia: Systematic Review and Meta-analysis",
journal="J Med Internet Res",
year="2021",
month="Aug",
day="27",
volume="23",
number="8",
pages="e28974",
issn="1438-8871",
doi="10.2196/28974",
url="https://www.jmir.org/2021/8/e28974",
url="https://doi.org/10.2196/28974",
url="http://www.ncbi.nlm.nih.gov/pubmed/34448706"
}

@Manual{refundR,
    title = {refund: Regression with Functional Data},
    author = {Jeff Goldsmith and Fabian Scheipl and Lei Huang and Julia Wrobel and Chongzhi Di and Jonathan Gellar and Jaroslaw Harezlak and Mathew W. McLean and Bruce Swihart and Luo Xiao and Ciprian Crainiceanu and Philip T. Reiss and Erjia Cui},
    year = {2025},
    note = {R package version 0.1-38},
    url = {https://CRAN.R-project.org/package=refund},
    doi = {10.32614/CRAN.package.refund},
  }

@article{Wang2018,
  author    = {Wang, Xiqiang and Han, Dan and Li, Guoliang},
  title     = {Electrocardiographic manifestations in severe hypokalemia},
  journal   = {Journal of International Medical Research},
  year      = {2018},
  volume    = {48},
  number    = {1},
  pages     = {0300060518811058},
  doi       = {10.1177/0300060518811058},
  pmid      = {30509119},
  pmcid     = {PMC7287199},
  month     = {dec}
}

@article{Chen2026,
    author = {Chen, Sida and Barrett, Jessica K and Palma, Marco and Pan, Jianxin and Tom, Brian D M},
    title = {Bayesian semiparametric modeling of biomarker variability in joint models},
    journal = {Biometrics},
    volume = {82},
    number = {3},
    pages = {ujag146},
    year = {2026},
    month = {09},
    issn = {1541-0420},
    doi = {10.1093/biomtc/ujag146},
    url = {https://doi.org/10.1093/biomtc/ujag146},
    eprint = {https://academic.oup.com/biometrics/article-pdf/82/3/ujag146/70687106/ujag146.pdf},
}

@article{Mccarthy2023,
  title={From Meaningful Outcomes to Meaningful Change Thresholds: A Path to Progress for Establishing Digital Endpoints},
  author={Mc Carthy, Marie and Burrows, Kate and Griffiths, Pip and Black, Peter M. and Demanuele, Charmaine and Karlsson, Niklas and Buenconsejo, Joan and Patel, Nikunj and Chen, Wen-Hung and Cappelleri, Joseph C.},
  journal={Therapeutic Innovation \& Regulatory Science},
  volume={57},
  number={4},
  pages={706--715},
  year={2023},
  month={Jul},
  doi={10.1007/s43441-023-00502-8},
  url={https://doi.org/10.1007/s43441-023-00502-8},
  publisher={Springer}
}

@article{Sanchez2014,
  title={Functional principal component analysis as a new methodology for the analysis of the impact of two rehabilitation protocols in functional recovery after stroke},
  author={S{\'a}nchez-S{\'a}nchez, M Luz and Belda-Lois, Juan-Manuel and Mena-del Horno, Silvia and Viosca-Herrero, Enrique and Gisbert-Morant, Beatriz and Igual-Camacho, Celedonia and Bermejo-Bosch, Ignacio},
  journal={Journal of NeuroEngineering and Rehabilitation},
  volume={11},
  number={1},
  pages={134},
  year={2014},
  month={Sep},
  day={10},
  publisher={BioMed Central},
  doi={10.1186/1743-0003-11-134},
  url={https://doi.org/10.1186/1743-0003-11-134},
  issn={1743-0003}
}

@article{Zablocki2024,
  title={Using functional principal component analysis (FPCA) to quantify sitting patterns derived from wearable sensors},
  author={Zablocki, Rong W. and Hartman, Sheri J. and Di, Chongzhi and Zou, Jingjing and Carlson, Jordan A. and Hibbing, Paul R. and Rosenberg, Dori E. and Greenwood-Hickman, Mikael Anne and Dillon, Lindsay and LaCroix, Andrea Z. and Natarajan, Loki},
  journal={International Journal of Behavioral Nutrition and Physical Activity},
  volume={21},
  number={1},
  pages={48},
  year={2024},
  month={Apr},
  day={26},
  publisher={BioMed Central},
  doi={10.1186/s12966-024-01596-w},
  url={https://doi.org/10.1186/s12966-024-01596-w},
  issn={1479-5868}
}

@article{Thygesen2018,
  author    = {Thygesen, Kristian and Alpert, Joseph S. and Jaffe, Allan S. and Chaitman, Bernard R. and Bax, Jeroen J. and Morrow, David A. and White, Harvey D.},
  title     = {Fourth Universal Definition of Myocardial Infarction (2018)},
  journal   = {Journal of the American College of Cardiology},
  year      = {2018},
  volume    = {72},
  number    = {18},
  pages     = {2231--2264},
  doi       = {10.1016/j.jacc.2018.08.1038}
}

@article{Nijjer2010,
  author    = {Nijjer, S. S. and Burke, M. and Dahdal, M. T. and Dubrey, S. W.},
  title     = {Amyloid disease and the danger of late diagnosis},
  journal   = {BMJ Case Reports},
  year      = {2010},
  volume    = {2010},
  pages     = {bcr0220102767},
  doi       = {10.1136/bcr.02.2010.2767},
  pmcid     = {PMC3027836},
  month     = {oct}
}

@article{Henderson1975,
 ISSN = {0006341X, 15410420},
 URL = {http://www.jstor.org/stable/2529430},
 author = {C. R. Henderson},
 journal = {Biometrics},
 number = {2},
 pages = {423--447},
 publisher = {International Biometric Society},
 title = {Best Linear Unbiased Estimation and Prediction under a Selection Model},
 urldate = {2026-02-26},
 volume = {31},
 year = {1975}
}

@article{ShroutFleiss1979,
  title        = {Intraclass correlations: Uses in assessing rater reliability},
  author       = {Shrout, P. E. and Fleiss, J. L.},
  journal      = {Psychological Bulletin},
  year         = {1979},
  volume       = {86},
  number       = {2},
  pages        = {420--428},
  doi          = {10.1037/0033-2909.86.2.420},
  url          = {https://pubmed.ncbi.nlm.nih.gov/18839484/}
}

@article{McGrawWong1996,
  title        = {Forming inferences about some intraclass correlation coefficients},
  author       = {McGraw, Kenneth O. and Wong, S. P.},
  journal      = {Psychological Methods},
  year         = {1996},
  volume       = {1},
  number       = {1},
  pages        = {30--46},
  doi          = {10.1037/1082-989X.1.1.30},
  url          = {https://doi.org/10.1037/1082-989X.1.1.30}
}

@article{Tackney2024,
  title = {Unleashing the Full Potential of Digital Endpoints: Eight Questions that Need Attention},
  author = {Tackney, Mia S. and Carpenter, James R. and Villar, Sof\'{\i}a S.},
  journal = {BMC Medicine},
  year = {2024},
  volume = {22},
  pages = {413},
  doi = {10.1186/s12916-024-03590-x},
  url = {https://doi.org/10.1186/s12916-024-03590-x},
}

@article{Yang2024,
  title   = {Predicting health outcomes with intensive longitudinal data collected by mobile health devices: a functional principal component regression approach},
  author  = {Yang, Qing and Jiang, Meilin and Li, Cai and Luo, Sheng and Crowley, Matthew J. and Shaw, Ryan J.},
  journal = {BMC Medical Research Methodology},
  year    = {2024},
  volume  = {24},
  number  = {1},
  pages   = {69},
  doi     = {10.1186/s12874-024-02193-7},
  url     = {https://doi.org/10.1186/s12874-024-02193-7}
}

@misc{BoariCoelho2022,
  author       = {Boari Coelho, Daniel and Shida, Thiago Kenzo Fujioka and Costa, Thaisy Moraes and de Oliveira, Claudia Eunice Neves and de Castro Treza, Renata and Hondo, Sandy Mikie and Los Angeles, Emanuele and Bernardo, Claudionor and dos Santos de Oliveira, Luana and de Jesus Carvalho, Margarete},
  title        = {A dataset of overground walking full-body kinematics and kinetics in individuals with Parkinson’s disease},
  howpublished = {\url{https://figshare.com/articles/dataset/A_dataset_of_overground_walking_full-body_kinematics_and_kinetics_in_individuals_with_Parkinson_s_disease/14896881}},
  year         = {2022},
  note         = {Version 4, published on Figshare, CC BY 4.0. Accessed 2026},
}

@article{Shida2023,
  author  = {Shida, T. K. F. and Costa, T. M. and de Oliveira, C. E. N. and de Castro Treza, R. and Hondo, S. M. and Bernardo, C. and dos Santos de Oliveira, L. and de Jesus Carvalho, M. and Coelho, D. B.},
  title   = {A public data set of walking full-body kinematics and kinetics in individuals with {Parkinson}'s disease},
  journal = {Frontiers in Neuroscience},
  volume  = {17},
  year    = {2023},
  pages   = {992585},
}

@article{Cipriani2022,
  title        = {Low QRS Voltages in Cardiac Amyloidosis: Clinical Correlates and Prognostic Value},
  author       = {Cipriani, A. and De Michieli, L. and Porcari, A. and Licchelli, L. and Sinigiani, G. and Tini, G. and others},
  journal      = {JACC: CardioOncology},
  year         = {2022},
  volume       = {4},
  number       = {4},
  pages        = {458--470},
  doi          = {10.1016/j.jaccao.2022.08.007},
  pmcid        = {PMC9700257},
  url          = {https://www.ncbi.nlm.nih.gov/pmc/articles/PMC9700257/}
}

@article{Censi2016,
  title        = {P-wave variability and atrial fibrillation},
  author       = {Censi, Federica and Corazza, Ivan and Reggiani, Elisa and Calcagnini, Giovanni and Mattei, Eugenio and Triventi, Michele and Boriani, Giuseppe and others},
  journal      = {Scientific Reports},
  volume       = {6},
  pages        = {26799},
  year         = {2016},
  doi          = {10.1038/srep26799},
  url          = {https://www.nature.com/articles/srep26799}
}

@misc{FoodandDrugAdministration2023,
author = {{Food and Drug Administration}},
title = {{Digital Health Center of Excellence}},
url = {https://www.fda.gov/medical-devices/digital-health-center-excellence},
year = {2023}
}

@article{Goldsack2020,
author = {Goldsack, Jennifer C. and Coravos, Andrea and Bakker, Jessie P. and Bent, Brinnae and Dowling, Ariel V. and Fitzer-Attas, Cheryl and Godfrey, Alan and Godino, Job G. and Gujar, Ninad and Izmailova, Elena and Manta, Christine and Peterson, Barry and Vandendriessche, Benjamin and Wood, William A. and Wang, Ke Will and Dunn, Jessilyn},
doi = {10.1038/s41746-020-0260-4},
isbn = {4174602002604},
issn = {23986352},
journal = {npj Digital Medicine},
number = {1},
publisher = {Springer US},
title = {{Verification, analytical validation, and clinical validation (V3): the foundation of determining fit-for-purpose for Biometric Monitoring Technologies (BioMeTs)}},
url = {http://dx.doi.org/10.1038/s41746-020-0260-4},
volume = {3},
year = {2020}
}

@Manual{R,
     title = {R: A Language and Environment for Statistical Computing},
     author = {{R Core Team}},
     organization = {R Foundation for Statistical Computing},
     address = {Vienna, Austria},
     year = {2021},
     url = {https://www.R-project.org/},
}

@misc{Bakker2024,
   author = {Jesse P Bakker and Roland Barge and Bryan Cobb and Chas Cota and Christine C Guo and Bert Hartog and Nathalie Horowicz-Mehler and Elena S Izmailova and Samantha McClenahan and Stéphane Motola and Smit Patel and Oana Paun and Marian Schoone and Emre Sezgin and Thomas Switzer and Animesh Tandon and Willem van den Brink and Srinivasan Vairavan and Benjamin Vandendriessche and Bernard Vrijens and Jennifer C Goldsack},
   title = {V3+: An extension to the V3 framework to ensure user-centricity and scalability of sensor-based digital health technologies},
   url = {https://datacc.dimesociety.org/resources/v3-an-extension-to-the-v3-framework-to-ensure-user-centricity-and-scalability-of-sensor-based-digital-health-technologies/},
   year = {2024},
}

@article{Landers2021,
author = {Landers, Matthew and Dorsey, Ray and Saria, Suchi},
doi = {10.1159/000517885},
issn = {2504110X},
journal = {Digital Biomarkers},
number = {3},
pages = {216--223},
title = {{Digital Endpoints: Definition, Benefits, and Current Barriers in Accelerating Development and Adoption}},
volume = {5},
year = {2021}
}

@article{Servais2021,
author = {Servais, Laurent and Camino, Eric and Clement, Aude and McDonald, Craig M. and Lukawy, Jacek and Lowes, Linda P. and Eggenspieler, Damien and Cerreta, Francesca and Strijbos, Paul},
doi = {10.1159/000517411},
issn = {2504110X},
journal = {Digital Biomarkers},
number = {2},
pages = {183--190},
title = {{First Regulatory Qualification of a Novel Digital Endpoint in Duchenne Muscular Dystrophy: A Multi-Stakeholder Perspective on the Impact for Patients and for Drug Development in Neuromuscular Diseases}},
volume = {5},
year = {2021}
}

@misc{Colloud2023,
   author = {Seya Colloud and Thomas Metcalfe and Scott Askin and Shibeshih Belachew and Johannes Ammann and Ernst Bos and Timothy Kilchenmann and Paul Strijbos and Damien Eggenspieler and Laurent Servais and Chloé Garay and Athanasios Konstantakopoulos and Armin Ritzhaupt and Thorsten Vetter and Claudia Vincenzi and Francesca Cerreta},
   doi = {10.1038/s41746-023-00790-2},
   issn = {23986352},
   issue = {1},
   journal = { npj Digit. Med},
   month = {12},
   publisher = {Nature Research},
   title = {Evolving regulatory perspectives on digital health technologies for medicinal product development},
   volume = {6},
   year = {2023},
}

\end{document}